\documentclass[twocolumn,floatfix,superscriptaddress,longbibliography,pra,nobalancelastpage]{revtex4-2}

\usepackage{amssymb,amsmath,bm}
\usepackage{graphicx}
\usepackage{epstopdf}
\usepackage{color}
\usepackage{appendix}
\usepackage[colorlinks=true,citecolor=blue,linkcolor=magenta]{hyperref}
\usepackage{float}
\usepackage[normalem]{ulem}
\usepackage{bbold}
\usepackage{enumitem}
\usepackage[normalem]{ulem}
\usepackage{tabularray}
\usepackage[capitalize]{cleveref}
\usepackage{physics}
\usepackage{csquotes}
\usepackage{dsfont}
\usepackage{stmaryrd}
\usepackage{algorithm}
\usepackage[noend]{algpseudocode} 
\usepackage{booktabs}

\newcommand{\adj}[1]{\breve{#1}}          

\NewDocumentCommand{\diagram}{ O{1.0} O{1} O{./fig} m }{
  \vcenter{\hbox{\includegraphics[scale=#1,page=#2]{#3/#4.pdf}}}
}

\newcommand{\td}{\text{d}}
\newcommand{\bC}{\mathbb{C}}
\newcommand{\bR}{\mathbb{R}}

\newcommand{\cT}{\mathcal{T}}
\newcommand{\re}{\text{Re}}
\newcommand{\dsI}{\mathds{1}}
\DeclareMathOperator*{\argmin}{argmin}

\algrenewcommand\textproc[1]{\texttt{#1}}

\begin{document}

\title{Implicit differentiation of tensor network algorithms}

\author{Lander Burgelman}
\email{lander.burgelman@ugent.be}
\thanks{authors contributed equally.}
\affiliation{Department of Physics and Astronomy, Ghent University, Krijgslaan 299, 9000 Gent, Belgium}

\author{Anna Francuz}
\email{anna.elzbieta.francuz@univie.ac.at}
\thanks{authors contributed equally.}
\affiliation{University of Vienna, Faculty of Physics, Boltzmanngasse 5, 1090 Wien, Austria}

\author{Paul Brehmer}
\affiliation{University of Vienna, Faculty of Physics, Boltzmanngasse 5, 1090 Wien, Austria}

\author{Lukas Devos}
\affiliation{Center for Computational Quantum Physics, Flatiron Institute, New York, New York 10010, USA}

\author{\\Jutho Haegeman}
\affiliation{Department of Physics and Astronomy, Ghent University, Krijgslaan 299, 9000 Gent, Belgium}

\author{Frank Verstraete}
\affiliation{Department of Physics and Astronomy, Ghent University, Krijgslaan 299, 9000 Gent, Belgium}
\affiliation{Department of Applied Mathematics and Theoretical Physics, University of Cambridge, Wilberforce Road, Cambridge, CB3 0WA, United Kingdom}

\author{Bram Vanhecke}
\affiliation{University of Vienna, Faculty of Physics, Boltzmanngasse 5, 1090 Wien, Austria}

\begin{abstract}
The current leading approach to the variational optimization of projected entangled-pair states (PEPS) is based on automatic differentiation, which allows for a convenient evaluation of the energy gradient with respect to the local variational degrees of freedom.
However, evaluating the energy gradient not only remains a major computational bottleneck of the optimization procedure, but also suffers from frequent numerical instabilities.
In this work, we adopt recent advances in implicit differentiation techniques to address these challenges in PEPS optimization.
By reformulating the core step of the gradient computation in terms of a single characteristic equation for the contraction environment, we reduce the cost of the gradient computation and improve its scaling with the problem size.
By choosing a suitable parametrization of this characteristic equation based on the intrinsic symmetries of the contraction environment, we can directly remove instabilities from the global gradient computation that would otherwise arise from the derivatives of subroutines of the contraction algorithm.
Finally, we demonstrate how this approach drastically simplifies the practical implementation of stable gradient-based PEPS optimization.
\end{abstract}

\maketitle


\section{Introduction}

In recent decades, tensor networks have emerged as powerful computational \cite{Orus_review_2019,Banuls_review_2023} and analytical \cite{Cirac_review_2021} tools in the study of quantum and classical many-body systems.
Following the success of the density matrix renormalization group (DMRG) \cite{white_densitymatrix_1992,white_densitymatrix_1993} for one-dimensional systems, which is now understood as a variational method over the class of matrix product states (MPS) \cite{MPS_PBC,SCHOLLWOCK201196,Ostlund}, the ansatz was later extended to two dimensions with the invention of projected entangled-pair states (PEPS) \cite{verstraete_renormalization_2004}.
Both are now successfully used across many areas of physics, from strongly correlated systems and quantum information to high-energy physics, with applications ranging from ground states and finite-temperature simulations to excited states and real-time evolution.
Despite this broad applicability, PEPS methods remain significantly less utilized than MPS, owing to their higher computational complexity and challenges in their implementation.
Among the most common applications of PEPS is the ground-state optimization of infinite translationally invariant systems \cite{full_update,simple_update}, where the state of the art is gradient-based variational optimization \cite{Var_iPEPS1,Var_iPEPS2}.

The accurate computation of the energy gradient remains the main bottleneck in variational PEPS optimization, both in terms of computational cost and conceptual complexity. The energy itself can only be evaluated approximately, using an approximate contraction environment obtained from an iterative procedure such as the Corner Transfer Matrix Renormalization Group (CTMRG) \cite{nishino_corner_1996,orus_simulation_2009} or boundary matrix product state (boundary MPS) methods \cite{jordan_bMPS_2008,haegeman_diagonalizing_2017,zauner-stauber_variational_2018}. Consequently, when calculating the energy gradient, one needs to differentiate the procedure used to obtain this approximate environment. While early approaches relied on approximate schemes for the exact PEPS gradient \cite{Var_iPEPS1,Var_iPEPS2}, the adoption of automatic differentiation (AD) enabled the computation of exact gradients of  approximate PEPS contraction schemes \cite{liao_differentiable_2019}, which consequently led to a growing number of applications of these techniques to relevant physical systems \cite{Hasik_AFM_AD_2021,Ponsioen_AD_excitations_2022,Niu_CSL_2022,Zhang_VUMPS_AD_Kitaev_magnets_2023,He_triangularHubbard_2026,hu2025dynamicalspectralfunctionkagome,chen2026simulatingfermionicfractionalchern}. However, the direct application of AD tools leads to significant technical challenges that can be grouped into three categories: \emph{(i) efficiency}, \emph{(ii) stability} and \emph{(iii) ease of implementation}.

First, even when using a state-of-the-art fixed-point formulation \cite{christianson_reverse_1994,liao_differentiable_2019,naumann_introduction_2024} that avoids the memory overhead of backpropagating through all individual iterations of the contraction procedure, the presence of iterative subroutines gives rise to a \emph{nested} iterative procedure for the gradient computation that is computationally very costly.
Second, AD-based PEPS optimization routinely suffers from numerical instabilities arising from the differentiation of subroutines used in the contraction algorithm \cite{liao_differentiable_2019}.
Notably, these instabilities are purely artifacts of the differentiation procedure, whereas the actual energy gradient is entirely stable with respect to variations that cause them.
Despite progress in formally addressing these issues in specific cases \cite{francuz_stable_2025}, the lack of a general solution often leads to heuristic workarounds.
Finally, the conventional approach to AD-based PEPS optimization requires every primitive operation used in the energy evaluation to be compatible with an AD engine. This means that custom differentiation rules must be defined for every subroutine that is not natively supported -- a process that can be laborious and limiting in practice.

In this work, we adopt implicit differentiation techniques \cite{Fiacco1976,christianson_reverse_1994,blondel_efficient_2022} to address these three challenges. Instead of differentiating through each subroutine of the contraction algorithm, we reformulate the contraction result in terms of a single characteristic equation and differentiate only that.
This approach (i) reduces the cost and scaling of the gradient computation with problem size, in some cases even reducing it to a subleading step in the overall optimization procedure; (ii) removes numerical instabilities by exploiting the symmetries of the contraction environment to construct stable parameterizations of derivatives; and (iii) greatly simplifies the implementation by completely removing the need for differentiable subroutines, as the characteristic equation is formulated using only tensor network contractions.

In practice, this implicit differentiation approach ultimately reduces to a single core routine: solving a system of linear equations obtained from the automatic differentiation of the algebraic characteristic equations.
This core procedure is largely independent of the specific contraction algorithm.
Different contraction schemes can be used by simply substituting the appropriate characteristic equations and variable parameterizations.
Nevertheless, deriving the specific characteristic equations that can be solved efficiently and converge reliably when using iterative sparse linear solvers is highly non-trivial.
Therefore, in this manuscript, we provide explicit forms of characteristic equations for the most commonly used PEPS contraction algorithms.

The structure of this work is as follows.
We start in \cref{sec:peps_implicit_differentiation} by formulating the general problem of variational PEPS optimization using AD.
After introducing some notation and reviewing the well established fixed-point differentiation approach, we illustrate the core ideas behind our implicit differentiation approach in this general context.
We proceed by demonstrating this approach for the specific case of PEPS optimization using CTMRG in systems with a spatial $C_{4v}$ symmetry in \cref{sec:c4vctm}. Based on the insights gained from this example, we illustrate a similar procedure for the two most commonly used general contraction schemes in \cref{sec:boundarymps,sec:ctmrg}.
We benchmark the performance of our approach in these different settings in \cref{sec:benchmarks}, before concluding in \cref{sec:conclusion}.


\section{PEPS optimization and implicit differentiation}
\label{sec:peps_implicit_differentiation}

Given a quantum system defined on a translationally invariant square lattice with a local physical Hilbert space $P$ at each site, we can parametrize a state of this system as a PEPS by introducing a local rank-5 tensor
\begin{align}
    p 
    &: \diagram[1.0]{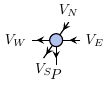},
\end{align}
as a multilinear map between the physical space $P$ and four virtual spaces $V_N, V_E, V_W, V_S$ corresponding to the four cardinal directions.
The dimension of the virtual spaces, called the bond dimension $D$, controls the amount of variational degrees of freedom contained in $p$, and by extension, dictates the expressiveness of the ansatz.
By tiling this tensor on an infinite square lattice and contracting all virtual indices, we can define a uniform PEPS obtained from $p$ as
\begin{equation}
    \ket{\psi(p)} = \diagram[1.0]{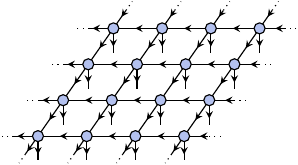}.
\end{equation}
While the arrows indicating the duality of the virtual spaces are important when working with (in particular, fermionic) symmetries, we will omit them from now on to lighten our notation.
Similarly, while a general uniform PEPS can be composed of a unit cell of different tensors, we will restrict our notation to the single-site case in the following.
All concepts and techniques can be generalized to the case of arbitrary unit cells.

\subsection{Variational PEPS optimization}
\label{sec:peps_optimization}

Given a Hamiltonian $H$, a primary application of the PEPS ansatz is to find the best approximation of its ground state by minimizing the variational energy
\begin{equation}
    \label{eq:variational_optimization}
    p = \argmin_p E(p) = \argmin_p \frac{\bra{\psi(p)} H \ket{\psi(p)}}{ \braket{\psi(p)}{\psi(p)}}.
\end{equation}
For a Hamiltonian $H$ defined as a sum of local terms $h$ that act on a few neighboring sites, 
evaluating the energy expectation value requires computing quantities of the form
\begin{equation}
    \bra{\psi(p)} h \ket{\psi(p)} = \diagram[1.0]{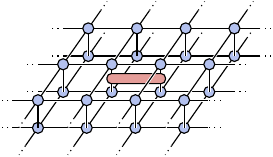}.
\end{equation}
To evaluate this expression, we need a reliable way to represent the infinite parts of the norm network $\braket{\psi(p)}{\psi(p)}$ in terms of an approximate \emph{contraction environment}
\begin{equation}
    \label{eq:peps_norm}
    \diagram[1.0]{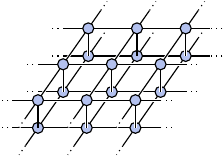}
    \mapsto
    \diagram[1.0]{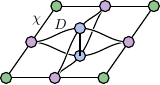}.
\end{equation}
This contraction environment, denoted as $x$ and represented by the purple and green tensors in \cref{eq:peps_norm}, implicitly depends on the variational degrees of freedom, $x \equiv x(p)$.
It provides an approximation of the infinite lattice, whose accuracy is controlled by the environment bond dimension $\chi$.
The two most common algorithms for obtaining such an environment are the corner transfer matrix renormalization group (CTMRG) \cite{nishino_corner_1996,orus_simulation_2009} and the boundary matrix product state (boundary MPS) \cite{haegeman_diagonalizing_2017,zauner-stauber_variational_2018} algorithms.
Both algorithms will be considered in this work.

Armed with an approximate contraction environment $x$, we can reformulate the variational problem for the total energy $E$ of \cref{eq:variational_optimization} in terms of an approximate energy density $e$, which takes the form 
\begin{align}
    &e(x, p) = \nonumber \\
    &\diagram[1.0]{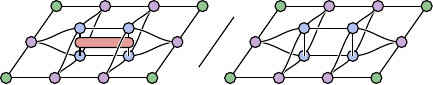}.
    \label{eq:peps_energy}
\end{align}

To optimize this approximate energy density, we can compute its gradient with respect to the variational parameters $p$ and use a gradient-based optimization scheme such as the quasi-Newton L-BFGS algorithm \cite{nocedal_numerical_2006}.
Using the chain rule, we can write the gradient as a sum of two terms corresponding to the explicit dependence of $e$ on $x$ and $p$ respectively,
\begin{equation}
    \label{eq:peps_gradient_chain_rule}
    \frac{\td e}{\td p}(p) = \partial_x e(x, p) \, \partial_p x + \partial_p e(x, p).
\end{equation}
The key step of a good PEPS algorithm is the practical evaluation of the first term on the right-hand side, which encodes the contribution to the total gradient arising from the dependence of the environment $x$ on the variational parameters $p$.
In this work, we will perform this key step using implicit differentiation.

Adopting some common notation in the context of AD, an introduction to which can be found in \cref{sec:ad_review}, we can rewrite \cref{eq:peps_gradient_chain_rule} in terms of the \emph{adjoint variables} corresponding to the contraction environment and variational parameters, defined as $\adj{x} = \partial_x e$ and $\adj{p} = \partial_p e$ respectively~\footnote{Here we have chosen to denote adjoints as $\adj{x}$ as opposed to the more conventional bar-notation $\bar{x}$, to avoid confusion with our notation for complex conjugation of tensors.}, which yields
\begin{equation}
    \label{eq:peps_gradient_chain_rule_bis}
    \frac{\td e}{\td p} = \adj{x} \, \partial_p x + \adj{p}.
\end{equation}
In this form, we see that the key operation in the gradient computation is the left action of the Jacobian $\partial_p x$ on the adjoint $\adj{x}$, known as a \emph{vector-Jacobian product} or VJP.
This VJP can be evaluated using a straightforward reverse-mode AD approach \cite{liao_differentiable_2019}, backpropagating through each primitive operation in the computational graph in reverse order.
This approach, however, requires a significant memory overhead and can suffer from instabilities of the primitive subroutines used in the algorithm, such as eigenvalue or singular value decompositions.

The key to reducing this memory overhead is to derive the VJP action of $\partial_p x$ based on an \emph{implicit} characterization of the contraction environment $x$.
We first review the simplest application of this idea, namely fixed-point differentiation \cite{liao_differentiable_2019,naumann_introduction_2024,francuz_stable_2025}, to introduce the necessary notation and highlight remaining difficulties with this approach.
We then proceed to the general formulation of implicit differentiation of a contraction algorithm in terms of set of purely algebraic characteristic equations.

For clarity, we restrict our notation in the main text to the setting of real variables. However, all results extend straightforwardly to the complex case.
A more rigorous treatment of real-valued cost functions of complex variables can be found in \cref{sec:ad_review}.

\subsection{Fixed-point differentiation}
\label{sec:fixed_point_differentiation}

A general contraction algorithm corresponds to an iterative procedure where we successively apply an iterating function $f$ starting from an initial guess $x_0$ for the environment, until convergence is reached.
In \emph{fixed-point} differentiation \cite{liao_differentiable_2019,naumann_introduction_2024,francuz_stable_2025}, we characterize the convergence of the contraction environment in terms of the \emph{fixed-point equation}
\begin{equation}
    \label{eq:fixed_point_equation}
    x = f(x, p),
\end{equation}
whose solution will be denoted as $x^*$.
Applying the implicit function theorem to this equation \cite{christianson_reverse_1994} by differentiating both sides at this solution yields
\begin{equation}
    \label{eq:fixed_point_implicit_derivative}
    \partial_p x = \partial_x f \, \partial_p x + \partial_p f.
\end{equation}
This allows us to write the VJP action of $\partial_p x$ as
\begin{equation}
    \label{eq:fixed_point_differentiation_vjp}
    \adj{x} \, \partial_p x = \adj{x} \left(\dsI - \partial_x f\right)^{-1} \, \partial_p f.
\end{equation}
Importantly, the VJP actions of $\partial_x f$ and $\partial_p f$ in this expression are evaluated only at the fixed-point solution $(x^*, p)$.
This means there is no need to keep track of any intermediate objects, solving the problem of the memory overhead encountered in straightforward reverse-mode AD.

To evaluate \cref{eq:fixed_point_differentiation_vjp} in practice, we can first compute an intermediate object $\adj{F}$ as the solution to the linear problem
\begin{equation}
    \label{eq:fixed_point_differentiation_linear_problem}
    \adj{F} \left( \partial_x f - \dsI \right) = \adj{x}.
\end{equation}
This system can be solved using any standard linear solver such as GMRES or BiCGSTAB.
Subsequently, we can plug this into the VJP action of $\partial_p f$ to obtain the desired result as
\begin{equation}
    \label{eq:fixed_point_differentiation_vjp_final}
    \adj{x} \, \partial_p x = - \adj{F} \, \partial_p f.
\end{equation}

It is important to note that in order to use fixed-point differentiation, it is necessary that the fixed-point equation \cref{eq:fixed_point_equation} is satisfied \emph{element-wise}.
This means that $f(x^*, p)$ produces tensors whose elements are all equal to those of $x^*$ within the convergence tolerance of the algorithm.
However, in practical implementations, \cref{eq:fixed_point_equation} is usually only satisfied up to allowed gauge transformations on the output of $f$.
This means that, in order to ensure element-wise convergence, any additional gauge freedom in the action of $f$ needs to be fixed explicitly using a suitable method \cite{francuz_stable_2025}.

While fixed-point differentiation resolves the memory overhead issue, it still requires the VJP actions of $\partial_x f$ and $\partial_p f$.
Obtaining these automatically via AD requires the function $f$ with all its subroutines to be fully differentiable.
In addition, the aforementioned instabilities in these primitive VJP actions can lead to instabilities in the overall gradient computation.
The fixed-point gradient linear problem \cref{eq:fixed_point_differentiation_linear_problem} is particularly susceptible to this.

Aside from these issues, a further motivation for going beyond fixed-point differentiation lies in the internal complexity of the iterating function $f$.
In most two-dimensional contraction schemes, $f$ is itself composed of iterative subroutines, such as sparse singular value decomposition or eigensolvers.
Crucially, evaluating the VJP corresponding to such an iterative subroutine requires solving a linear problem of the same nature as \cref{eq:fixed_point_differentiation_linear_problem}, and more importantly of a similar size.
Therefore, when applied to the specific setting of contraction algorithms with iterative subroutines, the core linear problem in fixed-point differentiation becomes \emph{nested}. Namely, for every function evaluation in the outer linear problem \cref{eq:fixed_point_differentiation_linear_problem}, we have to solve an inner linear problem of a similar size.

Extending the idea of implicit differentiation beyond the fixed-point condition of \cref{eq:fixed_point_equation} enables us to solve all of the remaining issues simultaneously.

\subsection{Implicit differentiation}
\label{sec:implicit_differentiation}

While characterizing the converged contraction environment $x^*$ as a fixed point of the iterating function $f$ is a natural choice, any implicit characterization could in principle be used.
In this framework, the solution to an optimization problem is implicitly defined as a root of an algebraic characteristic equation, which expresses the desired optimality conditions \cite{blondel_efficient_2022}.

This more abstract viewpoint allows us to \enquote{forget} about how $x^*$ was computed, and instead just focus on the conditions that characterize it.
In particular, if the optimality conditions characterizing $x^*$ are formulated as a single set of algebraic equations without iterative subroutines, the VJP action of $\partial_p x$ can in turn be evaluated in terms of a single un-nested linear problem.
In addition, the flexibility in choosing the characteristic equations allows us to exploit the symmetries of the contraction algorithm.
In particular, we can choose a parametrization of the equations specifically such that any instabilities that would arise from subroutines are completely removed from the global gradient computation.
Finally, to implement implicit differentiation in practice, we only need to be able to differentiate the characteristic equations themselves.
This is much easier than ensuring that the contraction algorithm as a whole is differentiable in a stable way.

Concretely, we will formulate implicit differentiation of a contraction algorithm in terms of a differentiable function $F(y, p)$ acting on a set of environment variables $y$ that implicitly depend on the variational parameters $p$, $y \equiv y(p)$.
The function $F$ is chosen such that the convergence of the environment corresponds to the solution of the \emph{characteristic equation}
\begin{equation}
    \label{eq:characteristic_equation}
    F(y, p) = 0.
\end{equation}
We will refer to the solution $y^*$ of this condition as the \emph{root} of the characteristic equation.
Here, we denote the variables representing the environment as $y$ rather than $x$, to emphasize that the characteristic equation can use a parametrization that is different from the one used in the iterating function.
In particular, this new set of variables $y$ may be expanded to include intermediate variables appearing in the iterating function, not only its final output $x$.
In practice, in the subsequent examples we will always introduce a simple and explicit map from the custom parametrization to the original environment variables $y \mapsto x$ from which we immediately obtain the corresponding map $\adj{x} \mapsto \adj{y}$ for the adjoints.
In particular, it will always be clear that the root $y^*$ of \cref{eq:characteristic_equation} is exactly equivalent to the fixed-point environment $x^*$.
In this formulation, the gradient computation \cref{eq:peps_gradient_chain_rule_bis} reduces to evaluating the VJP $\adj{y} \, \partial_p y$, since $\adj{x}\partial_p x = \adj{x}\partial_y x \partial_p y = \adj{y} \partial_p y$.

Differentiating the characteristic equation \cref{eq:characteristic_equation} at the root $y^*$ yields
\begin{equation}
    \label{eq:implicit_differentiation}
    \partial_{y} F \, \partial_p y = - \partial_p F.
\end{equation}
From this, we obtain the VJP action of $\partial_p x$ as
\begin{equation}
    \label{eq:implicit_differentiation_vjp}
    \adj{y} \, \partial_p y = \adj{y} \left(\partial_{y} F\right)^{-1} \left(-\partial_p F\right).
\end{equation}
Just as before, we can evaluate \cref{eq:implicit_differentiation_vjp} by computing an intermediate object $\adj{F}$ as the solution to the linear problem
\begin{equation}
    \label{eq:implicit_differentiation_linear_problem}
    \adj{F} \, \partial_{y} F = \adj{y},
\end{equation}
from which we can obtain the desired VJP action as
\begin{equation}
    \label{eq:implicit_differentiation_vjp_final}
    \adj{y} \, \partial_p y = - \adj{F} \, \partial_p F.
\end{equation}
In particular, this approach reduces to the fixed-point differentiation approach outlined before when we choose $y = x$ and $F(x, p) = f(x, p) - x$.

Algorithmically, the implicit differentiation approach can be summarized as follows.
First, the reverse-mode AD engine is called to backpropagate through the energy evaluation as described in \cref{alg:energy}.
While in the forward pass we contract the PEPS environment to obtain an energy value, in the reverse pass the contraction function is differentiated to accumulate the energy gradient.
Precisely in this reverse pass, in \cref{alg:contract}, we implement the implicit differentiation technique using \cref{eq:implicit_differentiation_linear_problem,eq:implicit_differentiation_vjp_final}.
We note that an explicit gauge-fixing step may need to be performed such that the optimality conditions $F(y, p)$ are indeed fulfilled at the fixed point, as will be pointed out in the forthcoming examples.

\begin{algorithm}[H]
    \caption{Energy evaluation and gradient}
    \label{alg:energy}
    \begin{algorithmic}[1]
    \Require PEPS tensor $p$, Hamiltonian $H$, initial environment $x_0$

    \Function{energy}{$p, x_0, H$}
    \State $p \leftarrow \texttt{symmetrize}(p)$ \Comment{if $C_{4v}$-symmetric} \label{line:symmetrization}
    \State $x \leftarrow \texttt{contract}(p, x_0)$ \Comment{Algorithm \ref{alg:contract}}
    \State \Return $e(p,x,H)$ \Comment{\cref{eq:peps_energy}}
    \EndFunction

    \Statex
    \State $E,\frac{\td e}{\td p} \leftarrow \texttt{backpropagate}(\texttt{energy}, p, x_0, H)$
    \end{algorithmic}
\end{algorithm}

\begin{algorithm}[H]
    \caption{PEPS contraction with a custom VJP}
    \label{alg:contract}
    \begin{algorithmic}[1]
    \Require PEPS tensor $p$, initial environment $x_0$, optimality conditions $F(y,p)$
    
    \Function{forward}{$p, x_0$}
    \State $x \leftarrow \texttt{contract}(p, x_0)$
    \State \Return $x$
    \EndFunction

    \Statex
    \Function{reverse}{$p$, $x$, $\adj{x}$}
    \State $y \leftarrow \texttt{parametrize}(x, \text{intermediates})$ \label{line:intermediates}
    \State $\adj{y} \leftarrow (\adj{x}\,\partial_y x, 0)$ \label{line:adjoint}
    \State $\adj{F} \leftarrow \texttt{solve}(\adj{F}\,\partial_y F = \adj{y})$ \Comment{\cref{eq:implicit_differentiation_linear_problem}}
    \State \Return $-\adj{F}\,\partial_p F$ \Comment{\cref{eq:implicit_differentiation_vjp_final}}
    \EndFunction
    \end{algorithmic}
\end{algorithm}

In the following sections, we will formulate a set of characteristic equations $F$ in a suitable parametrization $y$ for different contraction algorithms.
In this form they can be easily supplied into Algorithm \ref{alg:contract}, where the AD engine takes care of differentiating the algebraic expressions, and we use the result to solve a single linear problem.


\section{$C_{4v}$ CTMRG contraction}
\label{sec:c4vctm}

As a first guiding example, we illustrate the use of implicit differentiation for variational PEPS optimization using the CTMRG contraction algorithm in systems with a spatial $C_{4v}$ symmetry \cite{nishino_corner_1996}.
We start with a quick reminder of the $C_{4v}$ CTMRG algorithm. We then proceed to our main result and provide a set of characteristic equations in a suitable parametrization that can be used to apply implicit differentiation on top of any existing $C_{4v}$ CTMRG implementation.
In the remainder of this section, we will motivate this choice of parametrization and equations and discuss the improvements in performance and stability achieved by implicit differentiation.

\subsection{Contracting a PEPS using $C_{4v}$ CTMRG}
\label{sec:c4vctm_contraction}

At the level of an individual PEPS tensor, a $C_{4v}$ point group symmetry manifests itself as invariance under local rotation and Hermitian reflection,
\begin{equation}
    \label{eq:peps_tensor_symmetries}
    \diagram[1.0]{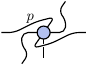} = \diagram[1.0]{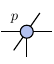},
    \quad
    \diagram[1.0]{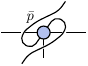} = \diagram[1.0]{peps_tensor_rhs},
\end{equation}
where $\bar{p}$ denotes the complex conjugation of the entries of $p$.
To ensure this spatial symmetry is preserved during optimizations, one can simply add an explicit projection of $p$ onto the symmetric subspace at the start of the energy computation (line~\ref{line:symmetrization} of Algorithm \ref{alg:energy}).

To simplify our notation, we will use a blocked single-layer tensor for the remainder of this work. In this notation, we express the PEPS norm and expectation value networks in terms of a local effective rank-4 tensor $T$, defined as the contraction of $p$ with its complex conjugate,
\begin{equation}
    \label{eq:peps_sandwich_blocked}
    \diagram[1.0]{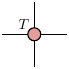} \equiv \diagram[1.0]{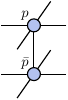}.
\end{equation}

The $C_{4v}$ CTMRG algorithm is a procedure for computing an environment $x = (C, E)$ consisting of a corner tensor $C$ and edge tensor $E$ that approximates the infinite parts of the norm network as depicted in \cref{eq:peps_norm}.
In the single-layer notation of \cref{eq:peps_sandwich_blocked}, the approximate contraction of $\braket{\psi(p)}{\psi(p)}$ takes the form
\begin{equation}
    \label{eq:peps_norm_c4v_ctmrg}
    \diagram[1.0]{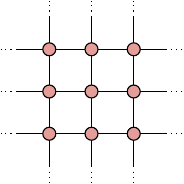}
    \mapsto
    \diagram[1.0]{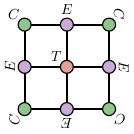}.
\end{equation}
This environment can be used to evaluate the variational energy in precisely the same manner as depicted in \cref{eq:peps_energy}.

The contraction algorithm itself is an iterative procedure, where every iteration maps $(C, E)$ to a new pair $(C', E')$ using an iterating function $f$ whose action $f((C, E), p) = (C', E')$ is defined as
\begin{align}
    \label{eq:c4vctm_enlarged_corner_eigh}
    \diagram[1.0]{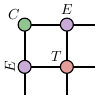} &\stackrel{\text{eigh}}{\approx} \lambda_C \diagram[1.0]{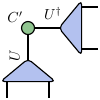}, \\
    \label{eq:c4vctm_edge_renormalization}
    \diagram[1.0]{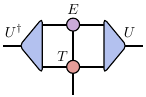} &= \lambda_E \diagram[1.0]{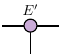}.
\end{align}
Here, the new corner $C'$ is obtained through a truncated Hermitian eigenvalue decomposition (\texttt{eigh}) of the enlarged corner matrix on the left hand side of \cref{eq:c4vctm_enlarged_corner_eigh}, and the scalars $\lambda_C$ and $\lambda_E$ correspond to the 2-norm of the left hand sides of \cref{eq:c4vctm_enlarged_corner_eigh,eq:c4vctm_edge_renormalization} respectively.

This procedure eventually produces a fixed-point environment $x^* = (C^*, E^*)$ for which $f((C^*, E^*), p) \approx (C^*, E^*)$ is reached up to some desired tolerance.
Note that, due to the gauge symmetry of the $C_{4v}$ contraction environment (which we will discuss in a broader context in \cref{sec:c4vctm_motivation}), achieving this kind of element-wise convergence in practice requires a suitable gauge-fixing of the environment tensors \cite{francuz_stable_2025}.
After gauge-fixing the environment tensors $(C^*, E^*)$, we use the same gauge transformation to gauge fix the fixed-point isometry, which we will denote as $U^*$.

\subsection{Implicit differentiation for $C_{4v}$ CTMRG}
\label{sec:c4vctm_implicit_differentiation}

In this section, we introduce the differentiable function $F(y, p)$ for which the root $y^*$ of the characteristic equation $F(y, p) = 0$ gives an implicit characterization of the fixed-point solution  $(C^*,E^*,U^*)$ of the $C_{4v}$ CTMRG algorithm.
Before doing so, we clarify the notation, in particular the input variables $y$ and the parametrization used to obtain a divergence free gradient.

As the set of differentiable variables, we choose $y = (C, E, u)$.
All other tensors that do not involve these variables are kept constant and are not differentiated. From now on, we depict these constant tensors in gray. 
The variables $C$ and $E$ directly represent the corner and edge tensors.
This choice is straightforward since the energy density depends explicitly on both tensors.
The variable $u$, on the other hand, corresponds to an \emph{intermediate} variable in line~\ref{line:intermediates} of \cref{alg:contract}, and arises from the parametrization of the isometry $U$.
While the energy density only depends on the intermediate variable $U$ implicitly through \cref{eq:c4vctm_enlarged_corner_eigh,eq:c4vctm_edge_renormalization}, we choose to include it as an explicit variable here.
Moreover, the fact that the energy density is not affected by any rotations within the relevant subspace, as shown in \cite{francuz_stable_2025} and discussed further in \cref{sec:c4vctm_motivation}, allows us to use a suitable parametrization of the isometry.

In particular, we choose to parametrize $U$ such that it only has variations along its null space projection $U_\perp$, defined as
\begin{equation}
   \label{eq:U_left_null}
    \diagram[1.0]{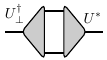} = 0.
\end{equation}
From this null space, we define a differentiable \enquote{projector} $U$ as
\begin{equation}
    \label{eq:ctm_U_parametrization}
    \diagram[1.0]{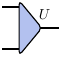}
    \equiv
    \diagram[1.0]{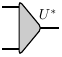}
    +
    \diagram[1.0]{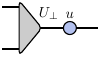}.
\end{equation}
In this parametrization, the tensors $U^*$ and $U_\perp^\dagger$ are treated as constants, meaning $U$ is only differentiable through its dependence on $u$.
Importantly, as we show in \cref{sec:c4vctm_motivation}, the choice of \cref{eq:ctm_U_parametrization} provides a proper divergence-free parametrization of the gradient of the isometry $U$.

Using this custom set of variables along with the definition \cref{eq:ctm_U_parametrization}, we can formulate a single set of characteristic equations $F((C, E, u), p) = 0$ that encodes the convergence criterion for $C$ and $E$, as well as the defining relations of the \texttt{eigh} decomposition in \cref{eq:c4vctm_enlarged_corner_eigh}. These equations take the form
\begin{equation}
    \label{eq:c4vctm_enlarged_corner_proj}
    \diagram[1.0]{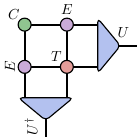} - \lambda_C \diagram[1.0]{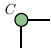} = 0,
\end{equation}
\begin{equation}
    \label{eq:c4vctm_edge_proj}
    \diagram[1.0]{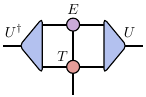} - \lambda_E \diagram[1.0]{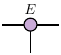} = 0,
\end{equation}
\begin{equation}
    \label{eq:c4vctm_enlarged_corner_proj_u}
    \diagram[1.0]{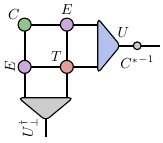} - \lambda_C \diagram[1.0]{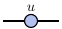} = 0,
\end{equation}
Here, we assume that $C$ and $E$ have unit normalization such that \cref{eq:c4vctm_enlarged_corner_proj,eq:c4vctm_edge_proj} hold with scalar factors $\lambda_C$ and $\lambda_E$, which are defined as the inner product of the first term in \cref{eq:c4vctm_enlarged_corner_proj} and \cref{eq:c4vctm_edge_proj} with $C$ and $E$ respectively.
Therefore, $\lambda_C$ and $\lambda_E$ are treated as differentiable variables, which depend on $y$.

The root $y^*$ of these characteristic equations exactly characterizes the $C_{4v}$ CTMRG fixed point.
Indeed, we clearly see that $y^* = (C^*, E^*, 0)$, where the fixed-point value $u^* = 0$ enforces the expected root value $U^*$ of the projector $U$ defined in \cref{eq:ctm_U_parametrization}.

Given this form of the characteristic equations $F$, we can directly apply the implicit differentiation approach of \crefrange{eq:implicit_differentiation}{eq:implicit_differentiation_vjp_final}.
Aside from the VJP action $\partial_y F$ evaluated at the root $y^*$, we also require the corresponding adjoint $\adj{y} = (\adj{C}, \adj{E}, \adj{u})$ as it appears on the right hand side of \cref{eq:implicit_differentiation_linear_problem} and line~\ref{line:adjoint} of \cref{alg:contract}.
The adjoints $\adj{x} = (\adj{C}, \adj{E})$ can be obtained automatically by backpropagating through the energy evaluation \cref{eq:peps_energy}, after which we can extract $\adj{y} = \adj{x}\partial_y x = (\adj{x}, 0)$.
This means that we simply set $\adj{u} = 0$, since the isometry $U$ is never used in the energy evaluation.

As the characteristic equation expressed by  \crefrange{eq:c4vctm_enlarged_corner_proj}{eq:c4vctm_enlarged_corner_proj_u} consists of a collection of tensor contractions, we can clearly see that the VJP action $\partial_p F$ also corresponds to a set of simple tensor contractions.
This means that implicit differentiation not only improves efficiency over conventional fixed-point differentiation, as we will argue below, but also that making $F$ differentiable is much easier in practice than making a truncated \texttt{eigh} decomposition differentiable in a stable manner.

As a final comment, we note some technical details that are important in implementing this form of implicit differentiation in practice.
First, to obtain the correct adjoint $\adj{C}$ to use in \cref{eq:implicit_differentiation_linear_problem}, we need to treat $C$ as a generic complex Hermitian matrix when backpropagating through the energy evaluation \cref{eq:peps_energy}.
This is necessary to allow its variations to be non-diagonal and complex, as is required to be consistent with the parametrization \cref{eq:ctm_U_parametrization}.
This naturally follows from the discussion below.
Importantly, this is only necessary in the reverse pass.
The forward contraction routine of \crefrange{eq:c4vctm_enlarged_corner_eigh}{eq:c4vctm_edge_proj} remains unchanged, and the fixed-point value $C^*$ is always real and diagonal.
Second, we have multiplied \cref{eq:c4vctm_enlarged_corner_proj_u} by the inverse of the fixed-point corner tensor ${C^*}^{-1}$, which serves as a constant preconditioner to improve the conditioning of the associated linear problem \cref{eq:implicit_differentiation_linear_problem}.
Equivalently, one can write these equations without ${C^*}^{-1}$ and instead supply its action as a right preconditioner to a linear solver.

\subsection{Motivating the characteristic equations}
\label{sec:c4vctm_motivation}

To understand the origin of the characteristic equations \crefrange{eq:c4vctm_enlarged_corner_proj}{eq:c4vctm_enlarged_corner_proj_u} and illustrate the benefits of the corresponding implicit differentiation, we now take a step back and analyze the fixed-point differentiation of $C_{4v}$ CTMRG.
We first pinpoint the issues this approach suffers from, and then explain how the implicit differentiation as presented above naturally resolves them.

Consider the fixed-point differentiation approach of \crefrange{eq:fixed_point_differentiation_vjp}{eq:fixed_point_differentiation_vjp_final} as applied to the iterating function defined in \cref{eq:c4vctm_enlarged_corner_eigh,eq:c4vctm_edge_renormalization}.
At increasing PEPS and environment bond dimensions, computing the truncated \texttt{eigh} decomposition in \cref{eq:c4vctm_enlarged_corner_eigh} from a full diagonalization quickly becomes infeasible, and one needs to resort to sparse iterative eigensolvers which build up the truncated subspace directly.
Without access to the full un-truncated decomposition, evaluating the VJP action corresponding to a truncated \texttt{eigh} requires solving a linear Sylvester equation \cite{francuz_stable_2025}.
This leads to a nested linear problem in \cref{eq:fixed_point_differentiation_linear_problem}, where every evaluation of the VJP action of $\partial_x f$ requires solving an inner linear Sylvester equation of a similar size.
This lies at the heart of the efficiency bottleneck.

In addition, evaluating the VJP action of a truncated \texttt{eigh} decomposition is inherently unstable when there are (nearly) degenerate eigenvalues.
Indeed, the stability of eigenvectors is related to the gaps in the spectrum, and they can exhibit non-analytic behavior in the case of degeneracies.
However, the energy evaluation in \cref{eq:peps_energy} is insensitive to rotations within subspaces of degenerate eigenvalues of \cref{eq:c4vctm_enlarged_corner_eigh}, or even with respect to general rotations in the kept subspace more generally.
It is easy to see that conjugating $C$ and $E$ by an arbitrary unitary $Q$, giving gauge-transformed corners and edges $(C', E')$, leaves all physical expectation values invariant,
\begin{equation}
    \label{eq:c4vctm_ctmrg_gauge_symmetry}
    \diagram[1.0]{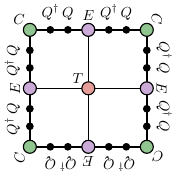} = \diagram[1.0]{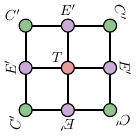}.
\end{equation}
This reparameterization corresponds to $U'  = U Q$, which is precisely a rotation of the basis in the kept subspace.
This rotation acts as a gauge freedom in the  $C_{4v}$ CTMRG algorithm, that is not visible to the \texttt{eigh} subroutine in isolation.
This gauge freedom implies that only the kept subspace is physically relevant, rather than a particular basis in which $C$ is diagonal.
The \texttt{eigh} subroutine is one convenient way of selecting the subspace associated with the dominant eigenvalues, but the particular basis of eigenvectors it selects within this subspace is irrelevant for the outer CTMRG algorithm.

We can resolve both the efficiency and stability issues through implicit differentiation.
By augmenting the usual system of fixed-point equations with equations that characterize the truncated subspace, parametrized in terms of a suitable set of variables $y = (C, E, u)$ as detailed in \cref{sec:c4vctm_implicit_differentiation}, we completely bypass the need to differentiate unstable subroutines such as \texttt{eigh}.
As the resulting characteristic equations $F(y, p) = 0$ are purely algebraic, we can compute the gradient from the un-nested linear problem \cref{eq:implicit_differentiation_linear_problem}, where evaluating the VJP action $\partial_{y} F$ corresponds to a set of simple and efficient tensor contractions.

To identify how this approach mitigates the instabilities mentioned above, we consider a general variation of the isometry $U$ around its fixed point value $U^*$,
\begin{equation}
    \label{eq:c4vctm_U_variation_general}
    \dot{U} = U^* \dot{\omega} + U_{\perp}\dot{u}.
\end{equation}
The contribution along $U^*$ satisfies the additional skew-hermiticity constraint $\dot{\omega} = -\dot{\omega}^\dagger$, while the contribution $\dot{u}$ along the left null space $U_\perp$ carries no further constraints.
It is well known that, when backpropagating through a truncated \texttt{eigh}, computing the adjoint associated with $\dot{\omega}$ requires dividing by differences of eigenvalues, which diverge in the case of degeneracies \cite{Matrix_derivatives,francuz_stable_2025}.
However, the skew-hermitian variations $\dot{\omega}$ exactly parameterize infinitesimal in-space rotations that can be compensated by the unitary gauge freedom $Q$, and can therefore be completely omitted. The only variations of $U$ that lead to variations in the energy are those parameterized by $\dot{u}$, associated with actual changes in the truncated subspace.
We stress that, in practice, there is no need to actually introduce a non-trivial $Q$ matrix at any stage of the $C_{4v}$ CTMRG algorithm.
Furthermore, the corner matrix $C$ as it appears in \cref{eq:c4vctm_enlarged_corner_eigh} always remains real and diagonal within the actual contraction routine.
The role of $Q$ is solely to exploit the gauge freedom in the gradient computation to enforce a stable parametrization of the variations.
We can therefore restrict $\dot{U}$ to the form
\begin{equation}
    \label{eq:c4vctm_U_variation_restricted}
    \dot{U} = U_\perp \dot{u},
\end{equation}
as long as we allow the variations of the corner tensor $\dot{C}$ to be non-diagonal and complex, but still Hermitian.
The difference between Eq.~\eqref{eq:c4vctm_U_variation_restricted} and Eq.~\eqref{eq:c4vctm_U_variation_general} is well known in the context of parameterizing the tangent space of the Grassmann versus the Stiefel manifold.
The adjoint associated with $\dot{u}$ will be determined by the spectral gap between the smallest kept eigenvalue and the largest truncated eigenvalue, and will be stable as long as we do not truncate through nearly degenerate eigenvalues. 

To impose the parametrization \cref{eq:c4vctm_U_variation_restricted} of the variations $\dot{U}$ in practice, we use a custom parametrization of the characteristic equations where we simply define a variable $U$ in terms of a single differentiable variable $u$ as in \cref{eq:ctm_U_parametrization}.
This $u$ vanishes at zeroth order, but its infinitesimal variations parameterize the variations of $\dot{U}$.
This also explains the earlier statement that the corner tensor should be treated as a generic complex Hermitian matrix when backpropagating through the energy evaluation.

Finally, we motivate our choice to treat the eigenvalues $\lambda_C$ and $\lambda_E$ occurring in \cref{eq:c4vctm_enlarged_corner_proj,eq:c4vctm_edge_proj} as differentiable quantities defined in terms of the variables $y$, rather than directly including them as independent variables in their own right.
Instead of assuming that $C$ and $E$ have unit normalization and then defining $\lambda_C$ and $\lambda_E$ as the inner product of the first term in \cref{eq:c4vctm_enlarged_corner_proj,eq:c4vctm_edge_proj} with $C$ and $E$ respectively, we could alternatively say that with these definitions of $\lambda_C$ and $\lambda_E$, \cref{eq:c4vctm_enlarged_corner_proj,eq:c4vctm_edge_proj} actually \emph{impose} the unit normalization of $C$ and $E$.
In practice, this strategy of making $\lambda_C$ and $\lambda_E$ depend on $y$ eliminates zero eigenvalues in the Jacobian $\partial_p F$ of these characteristic equations used in computing the VJP of the contraction routine.
This effect is explained in more detail for the simpler case of an eigenvalue problem in \cref{sec:differentiating_eigenvalue_problems}.
A similar approach of treating eigenvalues as differentiable quantities defined in terms of tensor variables will be used in formulating characteristic equations for other contraction algorithms below.

All these considerations eventually lead to the exact form of the characteristic equations given in \crefrange{eq:c4vctm_enlarged_corner_proj}{eq:c4vctm_enlarged_corner_proj_u}.
Note that, if we differentiate the characteristic equations \crefrange{eq:c4vctm_enlarged_corner_proj}{eq:c4vctm_enlarged_corner_proj_u} at the root $y^* = (C^*, E^*, 0)$, at first order we exactly recover the derivative of the deformed CTMRG procedure introduced in Ref.~\cite{francuz_stable_2025}, indicating that this form of $F$ indeed removes all divergences associated with degenerate eigenvalues.

\subsection{Implicit differentiation for different $C_{4v}$ contraction algorithms}
\label{sec:c4v_others}

To close this section, we briefly comment on how the implicit differentiation approach outlined above can be directly applied to different $C_{4v}$ symmetric contraction algorithms beyond CTMRG.
The symmetry conditions \cref{eq:peps_tensor_symmetries} ensure that the row-to-row transfer matrix corresponding to the PEPS norm network on the left-hand side of \cref{eq:peps_norm} is a Hermitian operator.
In this case, it is known that the contraction itself can be formulated as a variational problem, and that different methods for reaching this variational optimum give rise to equivalent contraction environments \cite{vanderstraeten_variational_2022}.
In particular, under these conditions, different contraction environments can be explicitly mapped to each other.
This means that, given a contraction environment obtained from any contraction algorithm compatible with $C_{4v}$ symmetry \cite{haegeman_diagonalizing_2017,nietner_efficient_2020,hauru_riemannian_2021,zhang_accelerating_2025a}, we can always map this environment to a set of parameters $y = (C, E, u)$ that satisfy the characteristic equations \crefrange{eq:c4vctm_enlarged_corner_proj}{eq:c4vctm_enlarged_corner_proj_u}.
Importantly, this implies that we can actually use the same implicit differentiation approach outlined above regardless of the specific contraction algorithm used in the forward pass.
Therefore, while we have focused on $C_{4v}$ CTMRG to motivate the form of the characteristic equations, our techniques can be easily incorporated into any existing code base without having to modify the implementation of the contraction algorithm itself.
On the other hand, the set of characteristic equations \crefrange{eq:c4vctm_enlarged_corner_proj}{eq:c4vctm_enlarged_corner_proj_u} is not unique.
Ref. \onlinecite{finite_entanglement_scaling} uses a different smaller set of equations which characterize a $C_{4v}$ symmetric contraction environment only in terms of the corner $C$ and isometry $U$.
One could use these equations to perform implicit differentiation by performing the energy evaluation \cref{eq:peps_energy} using the isometry $U$.
Details on how to map different symmetric contraction environments to each other and how the tensors within these environments are related to each other can be found in \cref{sec:symmetric_fixed_points}.


\section{Boundary MPS contraction}
\label{sec:boundarymps}

We further demonstrate the versatility of our approach by applying it to PEPS contractions using the boundary MPS method \cite{haegeman_diagonalizing_2017}.
In this approach, we view the PEPS norm network $\braket{\psi(t)}{\psi(t)}$ on the left hand side of \cref{eq:peps_norm} as an infinite power of a row-to-row transfer operator $\cT$, defined as
\begin{equation}
    \label{eq:boundarymps_transfer}
    \cT =
    \diagram[1.0]{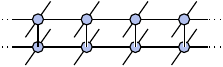}.
\end{equation}
To contract the norm network, we can then find the optimal MPS approximations to the dominant right and left eigenvectors of $\cT$,
\begin{align}
    \label{eq:boundarymps_top_fp}
    \diagram[1.0]{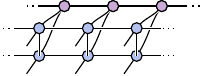}
    \propto
    \diagram[1.0]{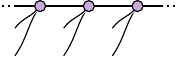}, \\
    \label{eq:boundarymps_bot_fp}
    \diagram[1.0]{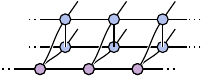}
    \propto
    \diagram[1.0]{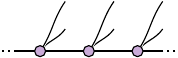},
\end{align}
which we will call the \emph{top} and \emph{bottom} MPS eigenvectors of $\cT$ respectively.

We start with a reminder of how these boundary MPS eigenvectors can be computed.
We then give a set of characteristic equations in a suitable parametrization that can be used to apply implicit differentiation on top of the resulting fixed point.
Just as before, this approach for computing the PEPS gradient can be directly applied on top of any existing implementation of boundary MPS contraction.
Additionally, the proposed characteristic equations have the advantage of removing the need for gauge fixing the resulting boundary MPS.

\subsection{Contracting a PEPS using boundary MPS}
\label{sec:boundarymps_contraction}

Given a transfer operator $\cT$ of the form \cref{eq:boundarymps_transfer}, one way to find its MPS eigenvectors is to use the variational uniform matrix product state (VUMPS) algorithm \cite{zauner-stauber_variational_2018}.
While we use this particular algorithm to guide our discussion, we stress that the implicit differentiation approach outlined below can be applied on top of any other algorithm that produces the same solution \cite{vanhecke_tangentspace_2021,hauru_riemannian_2021}.
We again adopt the blocked notation of \cref{eq:peps_sandwich_blocked}, but now without assuming any spatial symmetries of the PEPS tensor $p$.

When written in mixed canonical form \cite{vanhecke_tangentspace_2021}, the top eigenvector in \cref{eq:boundarymps_top_fp} can be defined in terms of a set of three variables $x^t = (A_L^t, A_R^t, C^t)$,
\begin{equation}
    \label{eq:boundarymps_mps_top}
    \diagram[1.0]{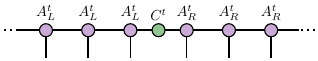}.
\end{equation}
The $A_L^t$ and $A_R^t$ tensors are left and right isometries, satisfying the conditions
\begin{equation}
    \label{eq:boundarymps_left_right_isometries}
    \diagram[1.0]{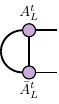}
    = \diagram[1.0]{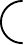},
    \qquad
    \diagram[1.0]{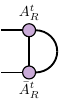}
    = \diagram[1.0]{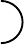}.
\end{equation}
In addition, the center-gauge bond tensor $C^t$ has the property that
\begin{equation}
    \label{eq:boundarymps_alc_car}
    \diagram[1.0]{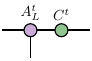} = \diagram[1.0]{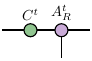},
\end{equation}
ensuring the MPS is translation invariant.

Starting from an initial boundary MPS $(A_L^t, A_R^t, C^t)$, a single iteration of the VUMPS algorithm produces an updated boundary MPS $( {A_L^t}', {A_R^t}', {C^t}' )$ as follows.
First, we find the left and right environments $G_L$ and $G_R$ as the leading eigenvectors of the left and right MPS transfer matrices respectively,
\begin{align}
    \label{eq:boundarymps_gl_top}
    \diagram[1.0]{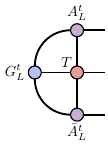}
    &= \lambda_L^t
    \diagram[1.0]{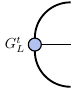},\\
    \label{eq:boundarymps_gr_top}
    \diagram[1.0]{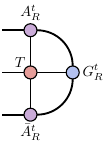}
    &= \lambda_R^t
    \diagram[1.0]{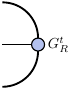}.
\end{align}
Second, we find updated \enquote{center} tensors ${A_C^t}'$ and ${C^t}'$ as the leading eigenvectors of the resulting one- and zero-site effective local Hamiltonians,
\begin{align}
    \label{eq:boundarymps_ac_update}
    \diagram[1.0]{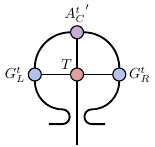}
    &= \lambda_{A_C}^t
    \diagram[1.0]{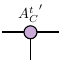},\\
    \label{eq:boundarymps_c_update}
    \diagram[1.0]{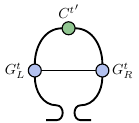}
    &= \lambda_C^t
    \diagram[1.0]{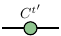}.
\end{align}
In practice, the eigenvalue problems for the extremal eigenvectors in \crefrange{eq:boundarymps_gl_top}{eq:boundarymps_c_update} are solved using a sparse Krylov-based eigensolver, such as the Arnoldi method.
To speed up convergence of the eigensolver subroutines, we can supply the tensors from the previous iteration as an initial guess for the corresponding eigenvectors.
Finally, we can use a suitable orthogonal factorization \cite{vanderstraeten_tangentspace_2019} to obtain updated isometries such that
\begin{equation}
    \label{eq:boundarymps_acc_update}
    \diagram[1.0]{boundarymps_ac_top_prime} = \diagram[1.0]{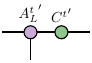} = \diagram[1.0]{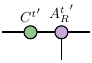}.
\end{equation}
If $\cT$ is a Hermitian operator, meaning the PEPS tensor $p$ has a Hermitian reflection symmetry, the bottom MPS eigenvector in \cref{eq:boundarymps_bot_fp} can be obtained simply as the hermitian conjugate of the top one.
In the case where $\cT$ does not have a reflection symmetry, the bottom MPS, parametrized in terms of variables $x^b = (A_L^b, A_R^b, C^b)$ as
\begin{equation}
    \label{eq:boundarymps_mps_bot}
    \diagram[1.0]{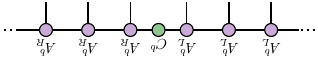},
\end{equation}
has to be computed separately.
This can be done using exactly the same procedure \crefrange{eq:boundarymps_gl_top}{eq:boundarymps_acc_update}, but where we rotate the transfer matrix $\cT$ by 180\textdegree.
Given both a top and bottom boundary MPS, evaluating the PEPS energy \cref{eq:peps_energy} additionally requires mixed environments $G_L$ and $G_R$ defined as
\begin{align}
    \label{eq:boundarymps_gl_mixed}
    \diagram[1.0]{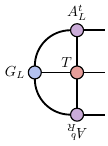}
    &= \lambda_L
    \diagram[1.0]{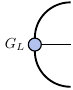},\\
    \label{eq:boundarymps_gr_mixed}
    \diagram[1.0]{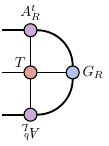}
    &= \lambda_R
    \diagram[1.0]{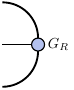}.
\end{align}
The relevant expressions for evaluating the PEPS energy density \cref{eq:peps_energy} using a boundary MPS contraction environment built from the tensors introduced here are given in \cref{sec:boundarymps_environments}.

We can define the iterating function $f$ corresponding to the VUMPS algorithm as a single combined update of the top and bottom boundary MPS tensors $x = (x^t, x^b) = (A_L^t, A_R^t, C^t, A_L^b, A_R^b, C^b)$. This $f$ produces updated tensors $({A_L^t}', {A_R^t}', {C^t}', {A_L^b}', {A_R^b}', {C^b}') = f((A_L^t, A_R^t, C^t, A_L^b, A_R^b, C^b), p)$ as summarized by \crefrange{eq:boundarymps_gl_top}{eq:boundarymps_acc_update} and their $(t \leftrightarrow b)$ analogues.
By repeatedly applying $f$, the procedure eventually converges to a set of fixed-point boundary MPS tensors $ x^*\equiv (A_L^{t*},A_R^{t*},C^{t*},A_L^{b*},A_R^{b*},C^{b*})$ that satisfy the fixed-point equation $x^* = f(x^*,p)$. Note that, at the fixed point, the eigenvalues in \crefrange{eq:boundarymps_gl_top}{eq:boundarymps_ac_update} all correspond to the same value $\lambda^t \equiv \lambda_L^t = \lambda_R^t = \lambda_{A_C}^t$, provided that $G^t_L$ and $G^t_R$ are normalized such that $\lambda^t_C=1$.
Importantly, for the implicit differentiation approach that follows, we only need this fixed-point condition to be valid only up to unitary gauge transformations.
This means that there is no need to ensure element-wise convergence using a gauge fixing procedure when using implicit differentiation here, in contrast to the case of fixed-point differentiation.
We denote the corresponding fixed-point values of the environment tensors \cref{eq:boundarymps_gl_top,eq:boundarymps_gr_top,eq:boundarymps_gl_mixed,eq:boundarymps_gr_mixed} as $G_L^{t*}$, $G_R^{t*}$, $G_L^{b*}$, $G_R^{b*}$, $G_L^{*}$ and $G_R^{*}$.

\subsection{Implicit differentiation for boundary MPS contraction}
\label{sec:boundarymps_implicit_differentiation}

To compute the gradient of the PEPS energy density obtained using a boundary MPS contraction environment, we again define a differentiable function $F(y, p)$ that implicitly characterizes the fixed-point solution $(A_L^{t*},A_R^{t*},C^{t*},A_L^{b*},A_R^{b*},C^{b*})$ as the root $y^*$ of the characteristic equation $F(y, p) = 0$.

For this we will use a custom set of 12 variables, \[y = (l^t, r^t, C^t, G_L^t, G_R^t, l^b, r^b, C^b, G_L^b, G_R^b, G_L, G_R).\]
The first five variables are used to characterize the top fixed-point MPS eigenvector, while the next five characterize the bottom fixed-point MPS eigenvector.
The last two variables are used to characterize the fixed-point mixed environments.
For each of these variables, we can write a corresponding characteristic equation.
In the language of \cref{alg:contract}, the variables $l^\alpha$, $r^\alpha$ and $C^\alpha$ directly relate to the variables $x$ of the iterating function $f$, while the others are intermediate variables that we have promoted to extend our set of equations beyond the simple fixed-point condition.

We start by considering the variables $y^t = (l^t, r^t, C^t, G_L^t, G_R^t)$ and the corresponding characteristic equations that characterize the top MPS eigenvector.
The first step is to use $l^t$ and $r^t$ to construct differentiable variables $A_L^t$ and $A_R^t$ in a way that ensures a stable parametrization of their variations.
Consider the constant left and right null spaces $V_L$ and $V_R$ of the fixed-point isometries $A_L^{t*}$ and $A_R^{t*}$ respectively, defined through
\begin{equation}
    \label{eq:boundarymps_vl_vr_top}
    \diagram[1.0]{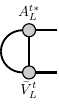} = 0,
    \qquad
    \diagram[1.0]{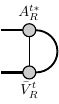} = 0.
\end{equation}
Using these null spaces, we define variables $A_L^t$ and $A_R^t$ as
\begin{align}
    \label{eq:boundarymps_al_top_parametrization}
    \diagram[1.0]{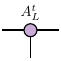}
    &= \diagram[1.0]{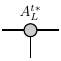} + \diagram[1.0]{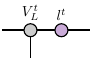},\\
    \label{eq:boundarymps_ar_top_parametrization}
    \diagram[1.0]{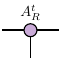}
    &= \diagram[1.0]{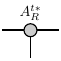} + \diagram[1.0]{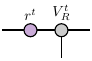}.
\end{align}
These variables are only differentiable through their dependence on $l^t$ and $r^t$ while all other tensors are kept constant.
This choice is motivated by the parameterization of the isometry in the $C_{4v}$ CTMRG case, and its validity is further discussed below.

Next, we define a center tensor variable $A_C^t$ as
\begin{equation}
    \label{eq:boundarymps_ac_top_parametrization}
    \diagram[1.0]{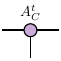} =
    \frac{1}{2}\left(
        \diagram[1.0]{boundarymps_al_c_top} + \diagram[1.0]{boundarymps_c_ar_top}
    \right)
\end{equation}
and a scalar $\lambda^t$ as
\begin{equation}
    \label{eq:boundarymps_lambda_top}
    \lambda^t = \diagram[1.0]{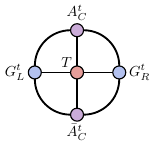}
\end{equation}
Using the definitions \crefrange{eq:boundarymps_al_top_parametrization}{eq:boundarymps_lambda_top}, we can formulate the first five characteristic equations $F^t(y^t, p)$ corresponding to the top fixed point as
\begin{gather}
    \label{eq:boundarymps_l_top_proj}
    \diagram[1.0]{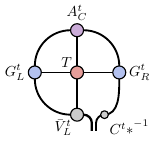} - \lambda^t \diagram[1.0]{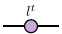} = 0,\\
    \label{eq:boundarymps_r_top_proj}
    \diagram[1.0]{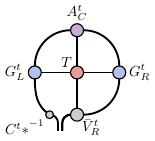} - \lambda^t \diagram[1.0]{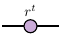} = 0.
\end{gather}
\begin{gather}
    \label{eq:boundarymps_c_top_proj}
    \frac{1}{2}
    \left(
    \diagram[1.0]{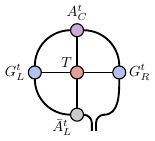} + \diagram[1.0]{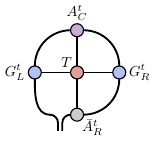}
    \right)
    - \lambda^t \diagram[1.0]{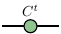} = 0,\\
    \label{eq:boundarymps_gl_top_bis}
    \diagram[1.0]{boundarymps_left_fp_apply_top} - \lambda^t \diagram[1.0]{boundarymps_left_fp_top} = 0,\\
    \label{eq:boundarymps_gr_top_bis}
    \diagram[1.0]{boundarymps_right_fp_apply_top} - \lambda^t \diagram[1.0]{boundarymps_right_fp_top} = 0,
\end{gather}
The root $y^{t*}$ of these top characteristic equations exactly corresponds to the top fixed-point boundary MPS $x^{t*} = (A_L^{t*}, A_R^{t*}, C^{t*})$ and its fixed-point environments $G_L^{t*}$ and $G_R^{t*}$.
Indeed, from \cref{sec:boundarymps_contraction} we clearly see that $y^{t*} = (0, 0, C^*, G_L^{t*}, G_R^{t*})$ satisfies $F^t(y^{t*}, p) = 0$. In particular, the root values $l^{t*} = 0$ and $r^{t*} = 0$ enforce the expected fixed-point values $A_L^{t*}$ and $A_R^{t*}$ of the variables \cref{eq:boundarymps_al_top_parametrization,eq:boundarymps_ar_top_parametrization}.

If $\cT$ has a Hermitian reflection symmetry, these five equations form a complete set of characteristic equations for the full contraction environment, as the bottom fixed point is simply the Hermitian conjugate of the top one.
Furthermore, if the system has a full $C_{4v}$ symmetry as in \cref{eq:peps_tensor_symmetries}, the fixed-point equations become exactly equivalent to the ones given in \cref{sec:c4vctm_implicit_differentiation}.
This is discussed in more detail in \cref{sec:symmetric_fixed_points}.
In the absence of spatial symmetries, we require five more equations for the variables $y^b = (l^b, r^b, C^b, G_L^b, G_R^b)$ that characterize the bottom MPS eigenvector.
These equations $F^b(y^b, p)$ are exact analogues of \crefrange{eq:boundarymps_vl_vr_top}{eq:boundarymps_gr_top_bis}, where we simply exchange the labels $t \leftrightarrow b$ and rotate $T$ by 180\textdegree.
Their root is given by $y^{b*} = (0, 0, C^{b*}, G_L^{b*}, G_R^{b*})$, which exactly corresponds to the bottom fixed-point MPS $x^{b*} = (A_L^{b*}, A_R^{b*}, C^{b*})$ and its fixed-point environment $G_L^{b*}$ and $G_R^{b*}$.

Finally, we require two more equations $F^g(y^g, p) = 0$ for the variables $y^g = (G_L, G_R)$, which are given by
\begin{gather}
    \diagram[1.0]{boundarymps_left_fp_apply_mixed} - \lambda \diagram[1.0]{boundarymps_left_fp} = 0,\\
    \diagram[1.0]{boundarymps_right_fp_apply_mixed} - \lambda \diagram[1.0]{boundarymps_right_fp} = 0.
\end{gather}
where $\lambda$ is defined as
\begin{equation}
    \lambda = \diagram[1.0]{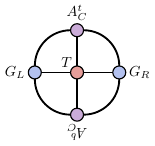}.
\end{equation}
Their root $y^{g*} = (G_L^*, G_R^*)$ simply corresponds to the fixed-point value of the mixed environments \cref{eq:boundarymps_gl_mixed,eq:boundarymps_gr_mixed}, provided that the fixed point tensors $G_L^\ast$ and $G_R^\ast$ are normalized such that 
\begin{equation}
    \diagram[1.0]{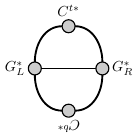} = 1.
\end{equation}

Altogether, this gives us a set of 12 characteristic equations $F(y, p) = 0$ for the 12 variables $y$ in the general case, whose root $y^*$ fully characterizes the boundary MPS contraction environment.
Given this form of $F$ and the root $y^*$, we can directly apply the implicit differentiation approach of \crefrange{eq:implicit_differentiation}{eq:implicit_differentiation_vjp_final} and supply it to \cref{alg:contract} to compute the PEPS gradient.

To solve the resulting linear problem \cref{eq:implicit_differentiation_linear_problem} in practice, we also need the adjoints $\adj{y}$ in line \ref{line:adjoint} of \cref{alg:contract}.
By backpropagating through the energy evaluation \cref{eq:peps_energy}, we obtain the adjoints of the boundary MPS tensors that appear explicitly in the contraction environment.
For variables that are not used in the energy evaluation, we simply set the corresponding adjoints to zero.
The precise form of the contraction environment and the corresponding adjoints is given in \cref{sec:boundarymps_environments}.
Finally, from the adjoints of the contraction environment we can obtain $\adj{y}$ by backpropagating through the parametrizations, $\adj{y} = \adj{x} \, \partial_y x$.
Specifically, from the adjoints $\adj{A}_L^\alpha$ and $\adj{A}_R^\alpha$, where $\alpha = t, b$, we obtain $\adj{l}^\alpha$ and $\adj{r}^\alpha$ by projecting onto the corresponding null spaces $V_L^\alpha$ and $V_R^\alpha$ respectively, according to \cref{eq:boundarymps_al_top_parametrization,eq:boundarymps_ar_top_parametrization}.

Note that, while we have found this particular formulation in terms of a single set of characteristic equations to be the most straightforward to implement and the most numerically stable in practice, alternative equivalent representations of the boundary MPS optimality conditions exist.
In particular, the problem can also be reformulated by decomposing it into smaller linear subproblems for the top and bottom boundaries separately, as described in \cref{sec:boundarymps_alternative}.

\subsection{Motivating the characteristic equations}
\label{sec:boundarymps_motivation}

To motivate this form of $F$ and the advantages it offers, we again start by considering how fixed-point differentiation using the iterating function $f$ defined in \cref{sec:boundarymps_contraction} works \cite{Zhang_VUMPS_AD_Kitaev_magnets_2023}.

To solve the fixed-point differentiation linear problem \cref{eq:fixed_point_differentiation_linear_problem} using a sparse solver, we need to evaluate the VJP action of $\partial_x f$ in every iteration.
Evaluating this VJP via automatic differentiation of a VUMPS iteration is highly non-trivial, as it requires that  every primitive operation within a single iteration, such as the QR decomposition used in \cref{eq:boundarymps_acc_update} and the dominant eigenvalue solves of \crefrange{eq:boundarymps_gl_top}{eq:boundarymps_c_update}, must be made differentiable while ensuring numerical stability at each step \cite{Zhang_VUMPS_AD_Kitaev_magnets_2023}.
Backpropagating through a dominant eigensolve requires solving a linear problem of a similar size as the original eigenvalue problem.
This means that the fixed-point differentiation of a boundary MPS contraction requires solving a nested linear problem for each dominant eigensolve inside the outer linear problem.

Implicit differentiation of the characteristic equation presented above circumvents these difficulties entirely.
On the one hand, the given form of $F$ requires the AD engine to differentiate only tensor contractions, completely removing the need to make the subroutines of a single iteration differentiable.
Furthermore, this means that the corresponding VJP action $\partial_{y} F$ also corresponds to a set of simple tensor contractions.
Therefore, rather than solving nested linear problems, by including the intermediate environment variables directly into the definition of $F$ we now only solve a single, albeit larger, linear problem for the adjoint $\adj{F}$.

To motivate our choice of variables $y$, and in particular the parametrization \cref{eq:boundarymps_al_top_parametrization,eq:boundarymps_ar_top_parametrization}, we consider the inherent symmetries of the boundary MPS approach.
Just as in the $C_{4v}$ CTMRG case, the boundary MPS contraction environment admits a gauge symmetry that can be exploited to restrict the variations around the fixed-point tensors to improve stability.
All physical observables and defining relations are invariant under the gauge transformation
\begin{align}
    \diagram[1.0]{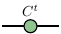}
    \mapsto
    \diagram[1.0]{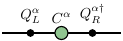},\\
    \diagram[1.0]{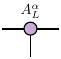}
    \mapsto
    \diagram[1.0]{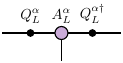},\\
    \diagram[1.0]{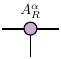}
    \mapsto
    \diagram[1.0]{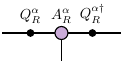},
\end{align}
where $Q_L^\alpha$ and $Q_R^\alpha$ are arbitrary unitary matrices.
In particular, this could be used to replace $C$ with a real diagonal matrix $S$ associated with its singular value decomposition $C = U S V^\dagger$, by choosing $Q_L= U^\dagger$ and $Q_R = V^\dagger$, although this is not a standard practice in the VUMPS algorithm.

In general, the variations around the fixed-point isometries, $\dot{A}_L^\alpha$ and $\dot{A}_R^\alpha$, contain contributions along both the primal solutions $A_L^{\alpha*}$ and $A_R^{\alpha*}$ respectively, as well as their corresponding null spaces, just as in \cref{eq:c4vctm_U_variation_general},
\begin{equation}
    \dot{A}_L^\alpha = A_L^{\alpha *} \dot{\omega}_L + V_L^{\alpha *}\dot{l}, \quad \dot{A}_R^\alpha = \dot{\omega}_R A_R^{\alpha *}  + \dot{r} V_R^{\alpha *},
\end{equation}
with again $\dot{\omega}_L$ and $\dot{\omega}_R$ skew-Hermitian.
However, we can use the gauge freedom to remove these contributions along $A_L^{\alpha*}$ and $A_R^{\alpha*}$ from the gradient computation entirely, while leaving the forward computation unchanged.
Even if we choose to include the step that makes $C$ real diagonal in the forward computation, it is then important to keep $C$ as a generic complex matrix in the parametrization of the reverse problem.
The restriction of the isometry variations is achieved through the parametrizations in \cref{eq:boundarymps_al_top_parametrization,eq:boundarymps_ar_top_parametrization} and their $(t \leftrightarrow b)$ equivalents, where $l^\alpha$ and $r^\alpha$ are used as differentiable variables in the characteristic equations.
This effectively restricts the variations of the isometries to only contain contributions along their null spaces, completely analogous to \cref{eq:c4vctm_U_variation_restricted}.


\section{CTMRG contraction}
\label{sec:ctmrg}

As a final example, we apply the principle of implicit differentiation to the asymmetric CTMRG algorithm for systems without spatial symmetries \cite{orus_simulation_2009}.
In this case, the contraction environment consists of four different corner tensors and four different edge tensors, as opposed to the single corner and edge tensor used in the $C_{4v}$ symmetric case of \cref{sec:c4vctm}.

We denote the edge and corner tensors as $\{C_\alpha\}$ and $\{E_\alpha\}$ respectively, where the index $\alpha$ labels the different directions.
We choose the convention that the directions $\alpha = 1, 2, 3, 4$ correspond to north, east, south and west for edges, and northwest, northeast, southeast and southwest for corners respectively.
This gives rise to an asymmetric contraction environment for the PEPS norm network of the form
\begin{equation}
    \label{eq:peps_nor_ctmrg}
    \diagram[1.0]{peps_norm_network_blocked}
    \mapsto
    \diagram[1.0]{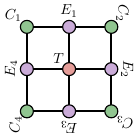}.
\end{equation}
We first briefly review how these corners and edges can be computed.
We then give a set of characteristic equations in a suitable parametrization that can be used to apply implicit differentiation on top of the resulting fixed point.
Finally, we motivate the form and parametrization of these characteristic equations by considering the symmetries of the CTMRG algorithm.

Throughout this entire section, we only explicitly provide diagrams centered around the $\alpha = 1$ direction, and implicitly assume analogous rotated expressions for the other directions $\alpha = 2, 3, 4$.
Just as before, we restrict our discussion here to the case of PEPS with a single-site unit cell.
Explicit expressions for the generalization to arbitrary unit cells are given in  \cref{sec:asymmetric_ctmrg_unit_cells}.

\subsection{Contracting a PEPS using CTMRG}
\label{sec:ctmrg_contraction}

Just as in the $C_{4v}$ symmetric case, the asymmetric CTMRG algorithm is an iterative procedure that maps a given contraction environment $x = (\{C_\alpha\}, \{E_\alpha\})$ consisting of a set of corner and edge tensors to a new set $(\{C'_\alpha\}, \{E'_\alpha\})$ by applying an iterating function $f$, $(\{C'_\alpha\}, \{E'_\alpha\}) = f((\{C_\alpha\}, \{E_\alpha\}), p)$.
In each iteration of $f$, we start by decomposing an enlarged environment using a truncated SVD as
\begin{equation}
    \label{eq:asymmctm_halfinf_env_svd}
    \diagram[1.0]{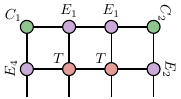} \stackrel{\text{SVD}}{\approx} \diagram[1.0]{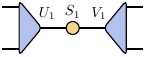},
\end{equation}
Denoting the inverses of the singular value matrices as $s_\alpha \equiv S_\alpha^{-1}$, we use the result of the decomposition to define left and right projectors for each direction as
\begin{align}
    \label{eq:asymmctm_pl}
    \diagram[1.0]{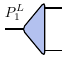} &= \diagram[1.0]{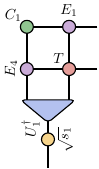},\\
    \label{eq:asymmctm_pr}
    \diagram[1.0]{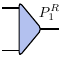} &= \diagram[1.0]{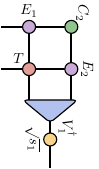}.
\end{align}
Using these projectors, we obtain the updated corners and edges through
\begin{align}
    \label{eq:asymmctm_corner_renormalization}
    \diagram[1.0]{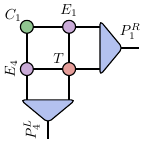} &= \lambda_{E_1} \diagram[1.0]{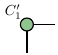},\\
    \label{eq:asymmctm_edge_renormalization}
    \diagram[1.0]{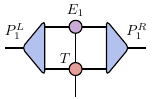} &= \lambda_{C_1} \diagram[1.0]{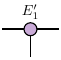},
\end{align}
and their corresponding rotated versions.
% ...
Here, the scalars $\lambda_{C_\alpha}$ and $\lambda_{E_\alpha}$ correspond to the 2-norm of the left hand sides of the respective equations, such that $\{C_\alpha\}$ and $\{E_\alpha\}$ have unit normalization at all times.
While we focus on the case of projectors $\{P_\alpha^L\}$ and $\{P_\alpha^R\}$ computed using the decomposition of a \enquote{half} infinite environment in \cref{eq:asymmctm_halfinf_env_svd}, the methods and results presented here also apply to other constructions.
In particular, the projectors could also be constructed from an SVD of a \enquote{full} infinite environment consisting of four enlarged corners \cite{Corboz_full_env_proj_2010}.

This procedure eventually produces a fixed-point environment $x^* = (\{C^*_\alpha\}, \{E^*_\alpha\})$ for which $f(x^*, p) \approx x^*$ is reached up to some desired tolerance.
Note that, again, ensuring this element-wise convergence requires a suitable gauge fixing of the fixed-point corner and edge tensors.
We then use this result to also gauge-fix the SVD tensors, where we denote the corresponding gauge-fixed fixed-point SVD tensors as $\{U^*_\alpha\}$, $\{S^*_\alpha\}$ and $\{V^*_\alpha\}$.

\subsection{Implicit differentiation for CTMRG}
\label{sec:ctmrg_implicit_differentiation}

The approach for formulating implicit differentiation for asymmetric CTMRG consists of the same steps as in the previous two examples.
Here, we use a custom set of variables $y$ defined as \[y = (\{\tilde{C}_\alpha\}, \{\tilde{E}_\alpha\}, \{u_\alpha\}, \{S_\alpha\}, \{v_\alpha\}).\]
The variables $\{\tilde{C}_\alpha\}$ and $\{\tilde{E}_\alpha\}$ correspond to \emph{modified} corner and edge tensors which are directly related to the variables $x = (\{C_\alpha\}, \{E_\alpha\})$ used in the energy evaluation.
The $\{u_\alpha\}$, $\{S_\alpha\}$, and $\{v_\alpha\}$ are the intermediate variables of line~\ref{line:intermediates} in \cref{alg:contract} and will be used to characterize the decomposition \cref{eq:asymmctm_halfinf_env_svd}.
The variables $S_{\alpha}$ are treated as generic complex matrices here, though they will reduce to the real diagonal matrices of singular values at the root of the characteristic equations.
For each of these variables, we will formulate a corresponding characteristic equation.

The first step is to use $\{u_\alpha\}$ and $\{v_\alpha\}$ to define differentiable tensors $\{U_\alpha\}$ and $\{V_\alpha\}$, whose derivatives lie only within the null spaces of the fixed-point isometries.
Consider the left and right null spaces $U_{\perp\alpha}$ and $V_{\perp\alpha}$ of $U^*_\alpha$ and $V^*_\alpha$ respectively, characterized as
\begin{equation}
    \label{eq:asymmctm_u_v_perp}
    \diagram[1.0]{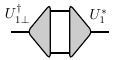} = 0,
    \qquad
    \diagram[1.0]{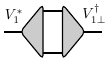} = 0.
\end{equation}
Using these constant null spaces, we define tensors $\{U_\alpha\}$ and $\{V_\alpha\}$ as
\begin{align}
    \label{eq:asymmctm_U_parametrization}
    \diagram[1.0]{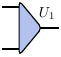} &= \diagram[1.0]{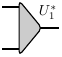} + \diagram[1.0]{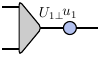},\\
    \label{eq:asymmctm_V_parametrization}
    \diagram[1.0]{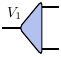} &= \diagram[1.0]{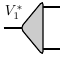} + \diagram[1.0]{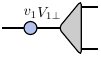},
\end{align}
which are differentiable only through their dependence on the variables $\{u_\alpha\}$ and $\{v_\alpha\}$.

We again denote the inverses of the variables $\{S_\alpha\}$ as $\{s_\alpha\}$, where $ s_\alpha = S_\alpha^{-1}$. Furthermore, we define the additional variables $s^L_\alpha$ and $s^R_\alpha$ as \enquote{modified square roots} of the inverse singular value matrices,
\begin{equation}
    \label{eq:asymmctm_sl_sr}
    s^L_\alpha = \sqrt[4]{s_\alpha^\dagger s_\alpha},
    \qquad
    s^R_\alpha = \sqrt[4]{s_\alpha s_\alpha^\dagger}.
\end{equation} 
From these basic variables, we define modified left and right projectors as
\begin{align}
    \label{eq:asymmctm_pl_tilde}
    \diagram[1.0]{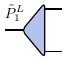} &= \diagram[1.0]{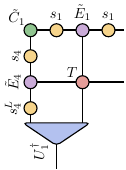},\\
    \label{eq:asymmctm_pr_tilde}
    \diagram[1.0]{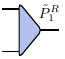} &= \diagram[1.0]{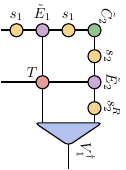}.
\end{align}

Using this custom set of variables and the above definitions, we can formulate a set of characteristic equations $F((\{\tilde{C}_\alpha\}, \{\tilde{E}_\alpha\}, \{u_\alpha\}, \{S_\alpha\}, \{v_\alpha\}), p) = 0$ as
\begin{gather}
    \label{eq:asymmctm_enlarged_corner_proj}
    \diagram[1.0]{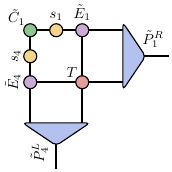} - \lambda_{C_1} \diagram[1.0]{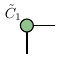} = 0,\\
    \label{eq:asymmctm_edge_proj}
    \diagram[1.0]{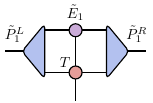} - \lambda_{E_1} \diagram{asymmctm_e_tilde} = 0,\\
    \label{eq:asymmctm_halfinf_env_proj_u}
    \diagram[1.0]{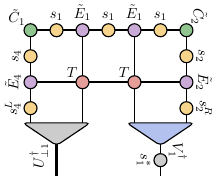} - \lambda_{S_1} \diagram[1.0]{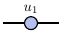} = 0,\\
    \label{eq:asymmctm_halfinf_env_proj}
    \diagram[1.0]{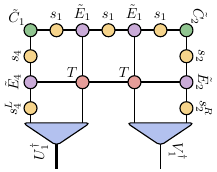} - \lambda_{S_1} \diagram[1.0]{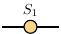} = 0,\\
    \label{eq:asymmctm_halfinf_env_proj_v}
    \diagram[1.0]{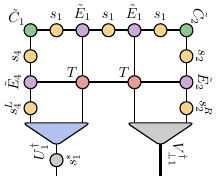} - \lambda_{S_1} \diagram[1.0]{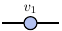} = 0.
\end{gather}
Here, we assume that $\{\tilde{C}_\alpha\}$ $\{\tilde{E}_\alpha\}$ have unit normalization, and have defined the scalars $\lambda_{C_1}$, $\lambda_{E_1}$ and $\lambda_{S_1}$ as the inner product of the first term in \cref{eq:asymmctm_enlarged_corner_proj,eq:asymmctm_edge_proj,eq:asymmctm_halfinf_env_proj} with $\tilde{C}_1$, $\tilde{E}_1$ and $S_1$ respectively.
This means that $\lambda_{C_1}$, $\lambda_{E_1}$ and $\lambda_{S_1}$ are also differentiable variables dependent on $y$.
Together with the rotated versions of \crefrange{eq:asymmctm_enlarged_corner_proj}{eq:asymmctm_halfinf_env_proj_v}, this gives us a set of 20 characteristic equations for the 20 variables $y$, which can be directly used to apply implicit differentiation \crefrange{eq:implicit_differentiation}{eq:implicit_differentiation_vjp_final} to compute the PEPS gradient.
Intuitively, the characteristic equations simultaneously encode the element-wise convergence of the corner and edge tensors, as well as the defining relations of the truncated SVD \cref{eq:asymmctm_halfinf_env_svd}.

To better understand the origin of our parametrization $y$, we start by considering the root $y^*$ of these equations.
From \cref{sec:ctmrg_contraction}, we can see that the root value is given by $y^* = (\{\tilde{C}_\alpha^*\}, \{\tilde{E}_\alpha^*\}, \{0\}, \{S_\alpha^*\}, \{0\})$ , where
\begin{align}
    \label{eq:asymmctm_c_e_tilde_fp}
    \diagram[1.0]{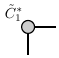} = \diagram[1.0]{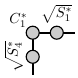},
    \qquad
    \diagram[1.0]{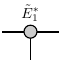} = \diagram[1.0]{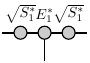},
\end{align}
First, note that $u_\alpha^*=0$ and $v_\alpha^*=0$ ensure the expected root values $\{U_\alpha^*\}$ and $\{V_\alpha^*\}$ of the isometries \cref{eq:asymmctm_U_parametrization,eq:asymmctm_V_parametrization}, meaning that $y^*$ indeed captures the fixed-point SVD tensors.
Second, we see that we can obtain the modified corner and edge tensors from the regular corners and edges by \enquote{pulling out} the square roots of the inverse singular value matrices that originally appeared in the projector definitions \cref{eq:asymmctm_pl,eq:asymmctm_pr}.
Indeed, this is exactly consistent with the modified projector definitions \cref{eq:asymmctm_pl_tilde,eq:asymmctm_pr_tilde} used in the characteristic equations.
This choice is motivated by the gauge symmetry of the CTMRG algorithm, which we will discuss in \cref{sec:asymmctm_motivation}.
We will see that square roots of singular value matrices do not transform covariantly, meaning that we have to treat them separately to ensure a stable parametrization of the isometries.

It is important to emphasize that, even though the characteristic equations make use of modified corners and edges, there is no need whatsoever to modify the actual CTMRG contraction algorithm to use implicit differentiation.
Indeed, to use \cref{alg:contract}, all we need is a way to relate the variables $y$ to the variables $x$ used in the regular contraction algorithm.
This relation can then be used to obtain the root $y^*$ and adjoint $\adj{y}$ from the output $x^*$ of the regular contraction algorithm and the adjoints $\adj{x}$.
Here, the relevant map $y \mapsto x$ is given by \cref{eq:asymmctm_U_parametrization,eq:asymmctm_V_parametrization} combined with the defining relation
\begin{align}
    \label{eq:asymmctm_c_e_tilde_to_c_e}
    \diagram[1.0]{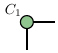} = \diagram[1.0]{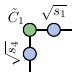},
    \qquad
    \diagram[1.0]{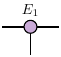} = \diagram[1.0]{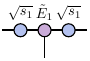}.
\end{align}
Indeed, given this definition we can directly obtain the root $y^*$ given above from the fixed-point environment $x^*$.
To obtain the adjoints $\adj{y}$, we first note that since $U_\alpha$ and $V_\alpha$ do not appear in the energy evaluation, their adjoints can simply be set to zero, giving $\adj{u}_\alpha = 0$ and $\adj{v}_\alpha = 0$.
Finally, from the adjoints $\adj{x}$ of the corner and edge tensors we can obtain adjoints of the variables $\tilde{C}_\alpha$, $\tilde{E}_\alpha$ and $\tilde{S}_\alpha$ by explicitly backpropagating through the defining relation \cref{eq:asymmctm_c_e_tilde_to_c_e} to complete $\adj{y}$.

\subsection{Motivating the characteristic equations}
\label{sec:asymmctm_motivation}

The reasoning behind the precise forms of $y$ and $F$ becomes clear when considering the subroutines and symmetries of the CTMRG algorithm.

The key step in the iterating function $f$ of the CTMRG contraction is the truncated SVD of \cref{eq:asymmctm_halfinf_env_svd}.
As a consequence, applying fixed-point differentiation to this $f$ gives a linear probem for the VJP action of $\partial_xf$ which involves backpropagating through this truncated SVD at each function application.
Without access to the full un-truncated decomposition, which is necessarily the case at large bond dimensions, backpropagating through a truncated SVD requires solving a linear Sylvester equation \cite{francuz_stable_2025}, again leading to a nested linear problem in \cref{eq:fixed_point_differentiation_linear_problem}.
In addition, the VJP action of a truncated SVD is inherently unstable when there are (nearly) degenerate singular values, while the energy evaluation \cref{eq:peps_energy} is entirely insensitive to these degeneracies, exactly as in the $C_{4v}$ symmetric case.

Just as before, the efficiency issue is solved by un-nesting the linear problem by treating the singular value tensors $\{U_\alpha\}$, $\{S_\alpha\}$ and $\{V_\alpha\}$ at the same level as $\{\tilde{C}_\alpha\}$, $\{\tilde{E}_\alpha\}$ and adding their defining equations directly to $F$.
To avoid the stability issues caused by degeneracies, we can again exploit a gauge symmetry of the contraction algorithm.
While this approach builds largely on the ideas from Section \ref{sec:c4vctm}, leveraging the gauge freedom is significantly more subtle in this case.

Just as we noted in \cref{sec:c4vctm_motivation}, in general the variations $\dot{U}_\alpha$ and $\dot{V}_\alpha$ of the truncated isometries appearing in \cref{eq:asymmctm_halfinf_env_svd} contain contributions along both their corresponding fixed-point values and their null space,
\begin{equation}
    \label{eq:asymmctm_U_V_variation_general}
    \dot{U}_\alpha = U_\alpha^* \dot{\omega}_L + U_{\perp\alpha} \dot{u}_\alpha,
    \quad
    \dot{V}_\alpha = \dot{\omega}_R V_\alpha^* + \dot{v}_\alpha V_{\perp\alpha}.
\end{equation}
Similar to the \texttt{eigh} case, the variations $\dot{\omega}_L$ and $\dot{\omega}_R$ lead to divergences in the adjoint of the enlarged environment when there are degenerate singular values \cite{Matrix_derivatives}.

However, just as in the $C_{4v}$ symmetric case, the CTMRG algorithm has a gauge symmetry that can be exploited to remove these contributions from the gradient computation.
Specifically, \cref{eq:asymmctm_U_V_variation_general} contains eight potentially diverging terms in total, one for each $U_\alpha$ and $V_\alpha$.
At the same time, given a CTMRG contraction environment, there are eight independent unitaries $Q_\alpha^L$ and $Q_\alpha^R$ which define a gauge transformation of this environment.
Namely, conjugating the corners and edges by these unitaries  leaves all physical observables invariant,
\begin{equation}
    \label{eq:asymmctm_gauge_symmetry}
    \diagram[1.0]{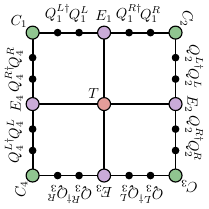} = \diagram[1.0]{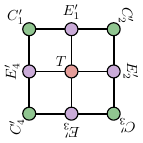}.
\end{equation}
We can interpret this gauge freedom of the environment as a freedom in redefining the left and right leading singular value subspaces,
\begin{equation}
    U_\alpha S_\alpha V_\alpha = U_\alpha Q_\alpha^{L\dagger} \left(Q_\alpha^L S_\alpha Q_\alpha^{R\dagger}\right) Q_\alpha^R V_\alpha.
\end{equation}
The gauge invariance of \cref{eq:asymmctm_gauge_symmetry} then means that, in principle, we only require projectors onto the leading subspaces rather than a specific basis in which $S_\alpha$ is diagonal.
If the entire CTMRG algorithm preserved this invariance, meaning that all intermediate objects transform appropriately under this set of gauge transformations, we could simply proceed in the same way as in \cref{sec:c4vctm_motivation} and exploit the gauge freedom to remove the problematic contributions from \cref{eq:asymmctm_U_V_variation_general}.

However, there are two issues that arise in the usual formulation of the CTMRG algorithm.
Firstly, the CTMRG projectors do not transform appropriately under gauge transformations due to the inverse square roots of the singular value matrices in \cref{eq:asymmctm_pl,eq:asymmctm_pr}.
The inverse square root $\sqrt{s_\alpha} \equiv \sqrt{S_\alpha^{-1}}$ transforms as
\begin{equation}
    \sqrt{s_\alpha} \rightarrow \sqrt{Q_\alpha^R S_\alpha^{-1} Q_\alpha^{L\dagger}} \neq Q_\alpha^R \sqrt{s_\alpha} Q_\alpha^{L\dagger}.
\end{equation}
Clearly, the final contracted environment used to compute observables always contains products of $\sqrt{s_\alpha}$, arising from pairs of left and right projectors, that yield an $s_\alpha$ with a proper transformation law.
However, the individual projectors do not transform covariantly.
Secondly, the edge tensor transformation $E_\alpha \to Q_\alpha^L E_\alpha Q_\alpha^{R\dagger}$ is clearly incompatible with the construction of environments consisting of multiple edge tensors whenever $Q_\alpha^L \neq Q_\alpha^R$, which are for example required when evaluating expectation values of operators acting on multiple sites.

Both issues can be solved using the following procedure.
The first logical step is to remove the inverse square roots from the projector definitions \cref{eq:asymmctm_pl,eq:asymmctm_pr} and include the $s_\alpha$ variables explicitly in the contraction environment.
This then also implies removing the inverse square roots from the renormalized corners and edges.
Altogether, this leads to the definitions of the modified corners, edges and projectors, $\{\tilde{C}_\alpha\}$, $\{\tilde{E}_\alpha\}$, $\{\tilde{P}_\alpha^L\}$ and $\{\tilde{P}_\alpha^R\}$, of \cref{eq:asymmctm_c_e_tilde_to_c_e,eq:asymmctm_pl_tilde,eq:asymmctm_pr_tilde}.
The appearance of inverse singular values $\{s_\alpha\}$ on the edges can be straightforwardly generalized to larger environments to systematically ensure gauge invariance.
However, the intermediate SVD step in \cref{eq:asymmctm_halfinf_env_svd} remains problematic.
The environment matrix of \cref{eq:asymmctm_halfinf_env_svd} is formed by cutting links between edge tensors $E_\alpha$.
In terms of the modified edge tensors $\{\tilde{E}_\alpha\}$, this leads to a dangling $\sqrt{s_\alpha}$ on outer legs of the edge tensors that are being cut.
This results in a non-unitary matrix transformation and consequently alters the SVD decomposition beyond a simple gauge transformation.
In a second step, we resolve this by replacing the dangling inverse roots on the left and right cuts by the custom roots $s_\alpha^L$ and $s_\alpha^R$ of \cref{eq:asymmctm_sl_sr}, with transformation laws
\begin{equation}
    s^L_\alpha \rightarrow Q_\alpha^L s_\alpha^L Q_\alpha^{L\dagger}, \qquad s^R_\alpha \rightarrow Q_\alpha^R s_\alpha^R Q_\alpha^{R\dagger}.
\end{equation}
This leads to the expressions for the infinite environment matrix on the left hand side of \crefrange{eq:asymmctm_halfinf_env_proj_u}{eq:asymmctm_halfinf_env_proj_v}, and ensures the consistency of the SVD decomposition.

After ensuring that observables calculated with CTMRG are invariant under unitary gauge transformations $Q_\alpha^R, Q_\alpha^L$, we can use their variations to cancel the diverging components in \cref{eq:asymmctm_U_V_variation_general}.
This restricts the isometry variations to their respective null spaces,
\begin{equation}
    \label{eq:asymmctm_U_V_variation_restricted}
    \dot{U}_\alpha = U_{\perp\alpha} \dot{u}_\alpha,
    \quad
    \dot{V}_\alpha = \dot{v}_\alpha V_{\perp\alpha}.
\end{equation}
In practice, this stable parametrization of the truncated isometry variations is easily imposed by using the parametrization of \cref{eq:asymmctm_U_parametrization,eq:asymmctm_V_parametrization} in the characteristic equations.
As a consequence, variations of singular value matrices $S_\alpha$ become \emph{general non-diagonal complex matrices}. 
This leads to a distinct operational treatment in the forward and backward passes:
\begin{itemize}
    \item \textbf{Forward pass:} All CTMRG computations remain exactly how they were described in \cref{sec:ctmrg_contraction} with real and diagonal $S_\alpha$ (\cref{alg:contract}, \texttt{forward} routine)
    \item \textbf{Reverse pass:} Throughout the reverse mode (\cref{alg:contract}, \texttt{reverse} routine) we treat $S_\alpha$ as complex non-diagonal matrices. This begins with defining the modified corners and edges in \cref{eq:asymmctm_c_e_tilde_to_c_e}.
\end{itemize}
Finally, we note that by examining the variations of \crefrange{eq:asymmctm_halfinf_env_proj_u}{eq:asymmctm_halfinf_env_proj_v} up to first order, we recover the expressions for the derivative of truncated SVD from \onlinecite{francuz_stable_2025}, but crucially without the divergent terms.


\section{Benchmarks}
\label{sec:benchmarks}

Finally, we benchmark the implicit differentiation approach presented above, against both the fixed-point differentiation scheme and a straightforward application of reverse-mode AD.
The benchmarks were carried out by extending the existing implementations of CTMRG and boundary MPS contractions in the PEPSKit.jl~\cite{brehmer_pepskit_2026} and MPSKit.jl~\cite{devos_mpskit_2026} software packages with implicit differentiation functionality.
The underlying tensor operations and symmetry capabilities are supported via TensorKit.jl~\cite{devos_tensorkitjl_2025}.
All implementations and benchmark data are publicly available in the open-source repository Ref.~\onlinecite{leburgel_implicit}.

\begin{figure}
    \centering
    \includegraphics{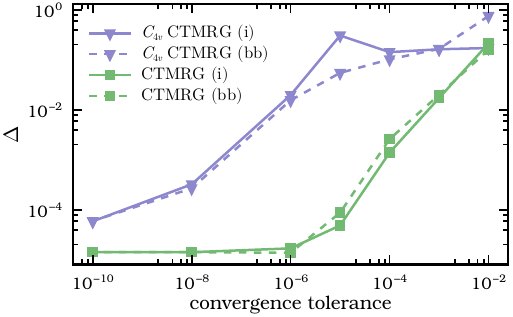}
    \caption{Relative distance $\Delta$ of the directional derivative of the PEPS energy density \cref{eq:peps_energy} computed with implicit differentiation (i) and with black-box reverse-mode differentiation (bb) as in \cref{eq:environment_jacobian_chain_rule_expanded}, to the value obtained using a symmetric first-order finite difference scheme with step size $\varepsilon = 10^{-3}$. The differences were averaged over three random PEPS points and directions with $D = 3$ and $\chi = 20$, and are shown as a function of the convergence tolerance of the contraction algorithm.}
    \label{fig:implicit_gradient_accuracy}
\end{figure}

We start by considering the accuracy of gradients obtained using the implicit differentiation approach compared to gradients resulting from straightforward \enquote{black-box} reverse-mode differentiation of~\cref{eq:environment_jacobian_chain_rule_expanded}.
For well-converged contraction environments, for which the characteristic equations \cref{eq:characteristic_equation} are satisfied up to high accuracy, we systematically observe that the gradients agree to within the specified linear solver tolerance.
In practice, however, it is not always possible or even desirable to fully converge the contraction environment.
Therefore, a natural question is how the gradient accuracy depends on the environment convergence.
To assess this, \cref{fig:implicit_gradient_accuracy} shows a comparison of the directional derivative of the PEPS energy density \cref{eq:peps_energy}, computed with both implicit differentiation and black-box reverse-mode differentiation, to a reference value computed using a simple first order finite-difference scheme.
We see that, as the environment is converged further, the gradients obtained via implicit differentiation approach the first order finite-difference gradients in the same way as those obtained through straightforward reverse-mode differentiation.
This means that even for relatively poorly converged environments, as long as we adapt the linear solver tolerance for \cref{eq:implicit_differentiation_linear_problem} accordingly, the resulting gradient is no less accurate than the one we would obtain with straightforward reverse-mode differentiation.
In particular, this level of accuracy may be perfectly adequate during the initial stages of the optimization.

Next, we benchmark the computational cost of gradient evaluation using implicit differentiation, as compared to a nested fixed-point differentiation approach, across a range of bond dimensions.
Three representative test cases will be considered, namely (i) $C_{4v}$ CTMRG, (ii) boundary MPS without spatial symmetries on a single-site unit cell, and (iii) CTMRG without spatial symmetries on a non-trivial unit cell.
For all test cases, we optimize a PEPS variationally using L-BFGS~\cite{nocedal_numerical_2006} at multiple values of the bond dimension $D$ and environment dimension $\chi$, storing intermediate checkpoints along the optimization trajectory.
For timing the contraction, the environment tensors are initialized using the environment from the previous L-BFGS iteration and are converged up to a tolerance of $10^{-7}$ in the relevant error measures.
The gradient times are obtained by measuring the time required to solve the linear system arising in the gradient computation.
In this way, we isolate the dominant computational cost in the reverse pass, while also reducing the effect of implementation details that might bias the benchmarks.
The gradient benchmark itself is performed using five moderately converged PEPS along the optimization trajectory, from which we take the median and standard deviation, once using the fixed-point formulation from~\cref{eq:fixed_point_differentiation_linear_problem} and once using the implicit differentiation approach of~\cref{eq:implicit_differentiation_linear_problem}.
In both cases, the resulting linear problems are solved using GMRES with a convergence tolerance of $10^{-6}$.
No explicit multi-threading is employed in our implementations.
Parallelism is restricted to the underlying BLAS routines, for which 8 CPU cores are allocated.
All benchmarks were performed on AMD EPYC 7713 compute nodes.

\subsection{$C_{4v}$ CTMRG}
\label{sec:benchmarks_c4vctmrg}

We first benchmark the case of $C_{4v}$-invariant CTMRG, for which we consider the spin-$1/2$ Heisenberg model
\begin{align}\label{eq:heisenberg_xyz}
    H = \sum_{\langle i,j\rangle} J_x S_i^x S_j^x + J_y S_i^y S_j^y + J_z S_i^z S_j^z,
\end{align}
where $S_i^\alpha$ denotes the spin-1/2 operator along the $\alpha$-axis on site $i$.
In order to capture the antiferromagnetic correlations of the ground state at $(J_x, J_y, J_z) = (1, 1, 1)$ in a single-site unit cell ansatz, we perform a unitary $-2iS^y$ sublattice rotation~\cite{Hasik_AFM_AD_2021}, amounting to a change in the parameters to $(J_x, J_y, J_z) = (-1, 1, -1)$.

\begin{figure}
    \centering
    \includegraphics{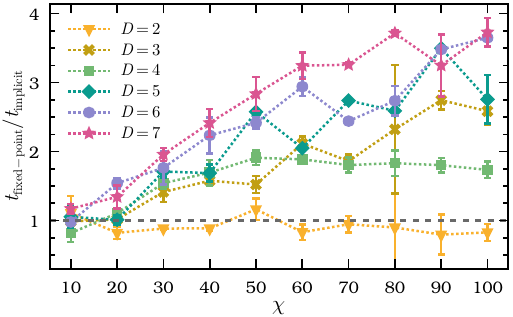}
    \caption{Speedup $t_\text{fixed-point} / t_\text{implicit}$ of the implicit differentiation approach over fixed-point differentiation, obtained from timing the gradient linear problem convergence for $C_{4v}$-invariant CTMRG on the Heisenberg model of \cref{eq:heisenberg_xyz}.}
    \label{fig:c4v_ctmrg_gradient_speedups}
\end{figure}

We evaluate the relative computational performance of the fixed-point and implicit differentiation approaches by computing the speedup $t_\text{fixed-point} / t_\text{implicit}$.
As shown in \cref{fig:c4v_ctmrg_gradient_speedups}, we consistently observe improved performance of implicit differentiation, with the speedup generally increasing with the total number of parameters.

\begin{figure}
    \centering
    \includegraphics{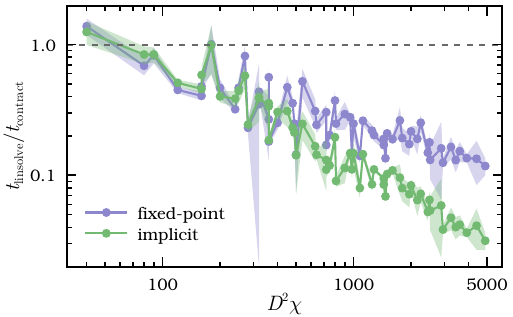}
    \caption{Gradient linear problem convergence times $t_\text{linsolve}$, normalized by the $C_{4v}$-invariant CTMRG contraction time $t_\text{contract}$, for both the implicit and fixed-point differentiation approaches on the Heisenberg model. We plot the data as a function of the matrix dimension of the $D^2\chi \times D^2\chi$-dimensional enlarged corner from \cref{eq:c4vctm_enlarged_corner_eigh} on a log-log scale.}
    \label{fig:c4v_ctmrg_gradient_times}
\end{figure}

When comparing the times $t_\text{linsolve}$ required to solve the linear problem of the gradient evaluation to the $C_{4v}$ CTMRG contraction times $t_\text{contract}$, we find that both differentiation approaches exhibit subleading computational complexity relative to the dominant contraction cost, as shown in~\cref{fig:c4v_ctmrg_gradient_times}.
Here we note that the CTMRG contraction steps are performed using a full \texttt{eigh} decomposition to obtain consistent scaling behavior.
That means that, since we do not utilize sparse eigensolvers, the CTMRG contraction cost may be particularly high at larger dimensions.
Nevertheless, the implicit differentiation method displays improved scaling with respect to the nested fixed-point approach, leading to increasing computational advantages with growing bond dimensions.

While we do observe overall increased performance, we note that the precise gradient timings are partially implementation-dependent.
Since the performance of the fixed-point gradient evaluation is strongly influenced by the specific realization of the pullback of the truncated \texttt{eigh} decomposition, we ensure conservative estimates by using a GMRES solver for solving the Sylvester equation therein
\footnote{Some implementations evaluate the Sylvester equation in the pullback using the LAPACK-based Bartels–Stewart algorithm~\cite{bartels_algorithm_1972}. However, significantly improved performance can be achieved already at moderate dimensions by solving the Sylvester equation using appropriately preconditioned Krylov-based iterative methods~\cite{francuz_stable_2025} such as GMRES.}.
Moreover, we observe that the contraction and gradient linear problem convergence dynamics may fluctuate across the optimization trajectory, making it difficult to control uniformly across values of $D$ and $\chi$.
We mitigate this effect by averaging the timings over several moderately optimized checkpoints.


\subsection{Single-site boundary MPS}

We proceed by benchmarking gradient computations for the boundary MPS contraction algorithm in the absence of spatial symmetries.
Specifically, we compare the implicit differentiation technique of \cref{sec:boundarymps_implicit_differentiation} with a fixed-point differentiation approach using the iterating function of \cref{sec:boundarymps_contraction}.
As before, we consider the Heisenberg model of \cref{eq:heisenberg_xyz} using a single-site unit-cell ansatz, and obtain the benchmarks following the same procedure as in the previous section.

\begin{figure}
    \centering
    \includegraphics{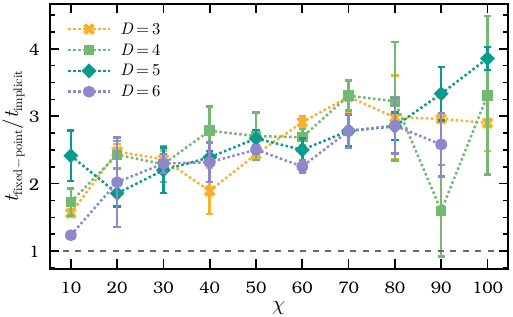}
    \caption{Speedup $t_\text{fixed-point} / t_\text{implicit}$ of the implicit over the fixed-point gradient linear problem, for boundary MPS without spatial symmetries on the sublattice-rotated Heisenberg model.}
    \label{fig:asymm_vumps_gradient_speedups}
\end{figure}

Considering the relative computational speedup $t_\text{fixed-point} / t_\text{implicit}$, we observe a clear speedup of implicit differentiation over the fixed-point approach, as shown in \cref{fig:asymm_vumps_gradient_speedups}.
The magnitude of the speedup is comparable to that observed for $C_{4v}$-symmetric CTMRG in \cref{fig:c4v_ctmrg_gradient_speedups} and again generally increases with the number of variational parameters.

\begin{figure}
    \centering
    \includegraphics{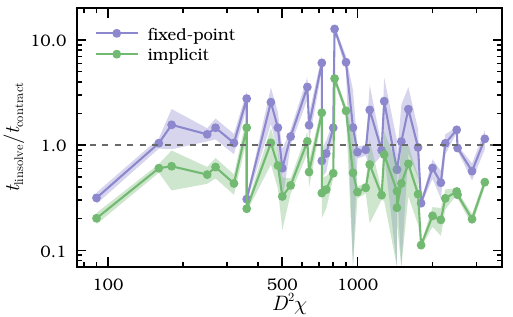}
    \caption{Times for solving the gradient linear problem, normalized by the boundary MPS contraction time, for both the fixed-point and implicit gradient on the Heisenberg model. The data is shown on a log-log scale as a function of the combined dimension $D^2\chi$ as before.}
    \label{fig:asymm_vumps_gradient_times}
\end{figure}

We also observe increased performance in the comparison between gradient linear solving times and contraction times shown in \cref{fig:asymm_vumps_gradient_times}.
In particular, as opposed to the fixed-point approach, the cost of the boundary MPS contraction generally outweighs the gradient computation in the implicit scheme.
We note, however, that the ratio between gradient and contraction times fluctuates substantially across the dimensions considered.
This can be partially attributed to the convergence dynamics of VUMPS-based boundary MPS contractions, which are typically more sensitive to the initial guess and may exhibit more erratic behavior along the optimization trajectory.


\subsection{CTMRG on non-trivial unit cells}

In the final benchmark setup, we consider CTMRG in the absence of spatial symmetries and with a non-trivial unit cell.
To this end, we study the Fermi--Hubbard model on the square lattice,
\begin{equation}
    H = -t\!\! \sum_{\langle i,j\rangle,\sigma} (
        \hat{c}^\dagger_{i\sigma} \hat{c}_{j\sigma} + h.c. 
    ) + U \sum_i \hat{n}_{i\uparrow} \hat{n}_{i\downarrow}
    - \mu \sum_i \hat{n}_i,
    \label{eq:fermi-hubbard}
\end{equation}
where $\hat{c}_{i\sigma}$ denotes the fermionic annihilation operator for spin $\sigma \in \{\uparrow, \downarrow\}$ at site $i$, $\hat{n}_{i\sigma}$ is the corresponding number operator, and $\hat{n}_i = \hat{n}_{i\uparrow} + \hat{n}_{i\downarrow}$.
The optimizations are performed at half-filling and $U/t=8$, using an iPEPS ansatz with a $2\times2$ unit cell in order to capture the antiferromagnetic correlations present in the model.
Furthermore, we impose an internal $f\mathbb{Z}_2 \times \mathrm{U}(1)$ symmetry on all tensors throughout the optimization and benchmarking procedures.
Using moderately optimized PEPS states, we perform the benchmarks following the same protocol as in the previous sections.

\begin{figure}
    \centering
    \includegraphics{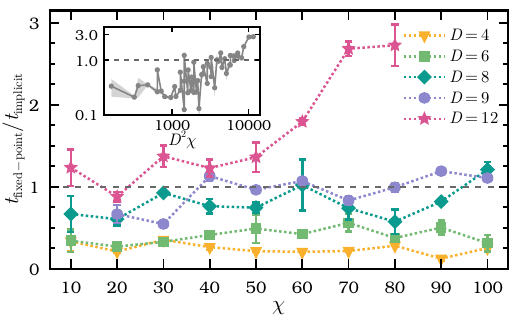}
    \caption{Ratio of the fixed-point and implicit gradient linear-solving times for CTMRG without spatial symmetries on the Fermi-Hubbard model. The inset shows the combined data on a log-log scale as a function of the total matrix dimension $D^2\chi$ of the half-infinite environment in \cref{eq:asymmctm_halfinf_env_svd}, indicating the asymptotic scaling in the total number of parameters.}
    \label{fig:asymm_ctmrg_hubbard_gradient_speedups}
\end{figure}

As before, we compare the computational performance of the fixed-point and implicit differentiation approaches by considering the ratio $t_\text{fixed-point} / t_\text{implicit}$ of their respective gradient linear-solving times.
The results are shown in \cref{fig:asymm_ctmrg_hubbard_gradient_speedups}.
In contrast to the previous benchmark scenarios, implicit differentiation does not consistently outperform the fixed-point approach.
A speedup only begins to emerge at comparatively large bond and environment dimensions, whereas for low and moderate dimensions the fixed-point method is typically faster.
Nevertheless, the ratio increases systematically with increasing $D$ and $\chi$, as indicated in the inset of \cref{fig:asymm_ctmrg_hubbard_gradient_speedups}, suggesting a more favorable asymptotic scaling of the implicit differentiation approach.

\begin{figure}
    \centering
    \includegraphics{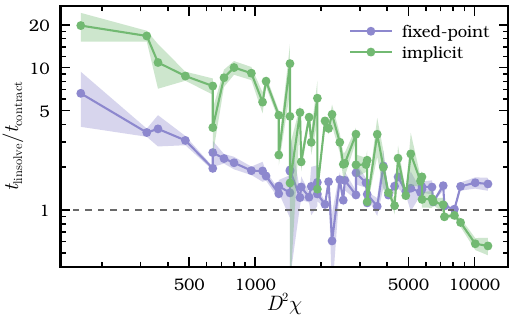}
    \caption{Fixed-point and implicit gradient linear-solving times, normalized by the CTMRG contraction time, on the Fermi-Hubbard model. The data is plotted on a log-log scale as a function of the combined dimension $D^2\chi$.}
    \label{fig:asymm_ctmrg_hubbard_gradient_times}
\end{figure}

To investigate this scaling behavior in more detail, we consider the normalized gradient linear-solving times shown in \cref{fig:asymm_ctmrg_hubbard_gradient_times}.
Indeed, we clearly observe that implicit differentiation exhibits an improved asymptotic scaling over the fixed-point approach.
While the timing curves intersect only at larger values of $D^2\chi$, suggesting a constant overhead in the implementation of the implicit approach, we do observe improved performance at higher dimensions.
Since in practical settings the differentiation techniques benchmarked here are employed together with sparse matrix decompositions that become performant only at high $D$ and $\chi$, the implicit scheme incurs a speedup in relevant scenarios with the computational cost dropping below that of the contraction.
We note that, as in \cref{sec:benchmarks_c4vctmrg}, conservative estimates are ensured by using a preconditioned iterative method to solve the coupled Sylvester equations appearing in the pullback of the truncated SVD that is present in the fixed-point CTMRG gradient.



\section{Conclusion}
\label{sec:conclusion}

In this manuscript, we have developed a new universal framework for differentiating tensor network algorithms.
It is based on implicit function differentiation and hence decouples the specific PEPS contraction algorithms from the differentiable functions that characterize their outputs.
Instead of differentiating the contraction subroutines directly, either via straightforward backpropagation or fixed-point differentiation, we differentiate sets of algebraic equations that fully characterize the output of these contraction algorithms.
This paradigm shift makes tensor network programming much simpler, more modular, and more stable, as it avoids the need to differentiate through often unstable subroutines.
Furthermore, as demonstrated by benchmarks, our method provides a measurable speed-up and improved asymptotic scaling over existing techniques while maintaining the ability to \enquote{warm-start} calculations by using initial guesses from previous iterations.

Although we have demonstrated this approach for three of the most common contraction schemes, the core idea presented in \cref{alg:energy} and \cref{alg:contract} is independent of the specific details and can be applied to other tensor network algorithms, or more generally, to any optimization problem in which the final result can be expressed in terms of differentiable optimality conditions.
Moreover, once the basic ingredients of \cref{alg:energy} and \cref{alg:contract} have been implemented for a given optimization problem, it is straightforward to experiment with different sets of algebraic characteristic equations to improve performance, modifying at most the input variables $y$ and their adjoints $\adj{y}$.

It is an interesting direction for future work to investigate whether more efficient or more compact sets of equations exist for the examples presented in this manuscript.
For instance, this includes alternative formulations for the $C_{4v}$ symmetric environment \cite{finite_entanglement_scaling}.
Although we have focused primarily on ground-state optimization, our method is applicable in a much broader context. In particular, it would be interesting to apply it to the excitation ansatz \cite{Vanderstraeten_PEPSexcitations_2019,Ponsioen_AD_excitations_2022}.


 \begin{acknowledgments}

We acknowledge inspiring discussions with Gleb Fedorovich, Wei Tang, Laurens Vanderstraeten, and Xingyu Zhang.
We are grateful to the authors of Ref.~\onlinecite{finite_entanglement_scaling} for useful discussions and coordinating our submissions.
L.B. and J.H.\ are supported by the European Union’s Horizon 2020 research and innovation programme through Grant No. 101125822 (ERC-CoG GaMaTeN).
P.B. is supported by the European Union’s Horizon 2020 research and innovation programme through Grant No. 863476 (ERC-CoG SEQUAM).
This research was funded in part by the Austrian Science Fund (FWF), Grant DOI: \href{https://doi.org/10.55776/ESP306}{10.55776/ESP306} (A.F.).
L.D.\ is supported by the Flatiron Institute.
The Flatiron Institute is a division of the Simons Foundation.
F.V.\ acknowledges funding from the UKRI grant EP/Z003342/1, BOFGOA (Grant No. BOF23/GOA/021), EOS (Grant No.~40007526) and IBOF (Grant No.~IBOF23/064).
Early stages of this work were performed in MATLAB with the help of the \texttt{ncon} function \cite{pfeifer_ncon_2015} for tensor contractions.
A.F. and B.V. are grateful to the Mathworks team for their assistance with certain undocumented aspects of the \texttt{dlfeval} functionality.
The computational results have been achieved using the Austrian Scientific Computing (ASC) infrastructure.

\end{acknowledgments}

\section*{Author Contributions}

F.V. conceptualized and implemented the initial idea of implicit differentiation of contraction algorithms based on a single set of algebraic characteristic equations in the context of $C_{4v}$ symmetric systems.
A.F. and B.V. formulated the characteristic equations for $C_{4v}$ CTMRG and boundary MPS contractions of \cref{sec:c4vctm_implicit_differentiation,sec:boundarymps_implicit_differentiation}, after which L.B. adapted this approach to formulate the characteristic equations for asymmetric CTMRG of \cref{sec:ctmrg_implicit_differentiation}.
L.B. and P.B. wrote the implementation of implicit differentiation for the different contraction schemes used in the benchmarks, which can be found in Ref.~\onlinecite{leburgel_implicit}.
J.H. and L.D. developed the framework for AD of (symmetric) tensor operations underlying all implementations of Ref.~\onlinecite{leburgel_implicit}, and wrote stable implementations of the differentiation rules for various numerical routines used in this work.
L.B. and A.F wrote the manuscript, while P.B. performed the numerical benchmarks.
All authors contributed to discussions, manuscript revisions, and approved the final version of the manuscript.


\appendix
\newpage


\section{Automatic differentiation and PEPS optimization}
\label{sec:ad_review}

In this Appendix we review the basic concepts of reverse-mode AD, and how it can be applied to the variational optimization of PEPS.

\subsection{Automatic differentiation in a nutshell}

Consider a set of variational parameters $p$, which we can interpret as an $n$-component complex vector $p \in \bC^n$.
A real-valued cost function $e$ of the complex parameters $p$ is given by a map $e : \bC^n \to \bR$.
We can take derivatives of such a cost function with respect to the complex variables $p$ using the formalism of Wirtinger derivatives.
Namely, for a complex scalar variable $z = x + i y$, we denote
\begin{equation}
\partial_z = \frac{1}{2}\left(\frac{\partial\ }{\partial x} - i \frac{\partial }{\partial y}\right),
\quad
\partial_{\bar{z}} = \frac{1}{2}\left(\frac{\partial\ }{\partial x} + i \frac{\partial }{\partial y}\right).
\end{equation}
If we consider a specific trajectory $p(t)$ of our variational degrees of freedom parametrized in terms of a single dummy variable $t$, we can use the chain rule to write the \emph{directional derivative} of the cost function as
\begin{equation}
    \label{eq:directional_derivative}
    \begin{split}
    \dot{e} &= \sum_k (\partial_{p^k}e) \dot{p}^k + (\partial_{\bar{p}^k}e) \dot{\bar{p}}^k \\
    &= 2 \,\re \left( \sum_k \partial_{p^k} e \, \dot{p}^k \right) = 2 \, \re \langle \adj{p}, \dot{p} \rangle.
    \end{split}
\end{equation}
Here, $\re \langle \adj{p}, \dot{p} \rangle \equiv \re \left(\adj{p}^\dagger \dot{p}\right)$ denotes the real inner product of two complex vectors, and we have relied on $\partial_{\bar{p}^k} e = \overline{\partial_{p^k} e}$ due to the fact that $e$ is real.
In this expression, $\dot{p}$ denotes the $n$-component complex vector $\dot{p}$ with entries given by $\dot{p}^k$, and will be referred to as the \emph{tangent vector}, or just \emph{tangent} for short.
Additionally, we have introduced $\adj{p}$ as a complex vector with entries given by $\partial_{\bar{p}^k} e$.
We will refer to $\adj{p}$ as the \emph{adjoint} of $p$, also called a \emph{cotangent vector} or \emph{cotangent} for short.

Now consider the case where our cost function computation is naturally interpreted as a two-stage process, where first we map our complex variables to some intermediate representation $g(p)$ using a function $g : \mathbb{C}^n \to \mathbb{C}^m$, and only then compute the cost function as $e(p) \equiv e(g(p))$, where now we interpret $e$ as a map $e : \mathbb{C}^m \to \mathbb{R}$. In this case, we can use the chain rule to find two equivalent expressions for \cref{eq:directional_derivative} as
\begin{equation}
    \label{eq:directional_derivative_bis}
    \dot{e} = 2 \re\langle \adj{g}, \dot{g} \rangle = 2 \re \langle \adj{p}, \dot{p} \rangle.
\end{equation}
Here we have introduced a tangent and adjoint for the intermediate variables $g$ in exactly the same way as before.

The basic idea of automatic differentiation is the efficient programmatic use of the chain-rule in evaluating the derivative of a composition of functions as in \cref{eq:directional_derivative_bis}.
The core idea becomes clear when we consider how to obtain the object we want, the total derivative $\dot{e}$, from the basic objects we have at our disposal, namely the tangent and adjoint that we always \enquote{know}.
In particular, we always know that the first tangent is $\dot{t} = \frac{\td t}{\td t} = 1$, and that the final adjoint is $\adj{e} = \partial_e e = 1$.
Considering the computational sequence $t \to p \to g \to e$, we could start from $\dot{t}$ and push the tangents forward through the sequence to obtain $\dot{e}$. This is called \emph{forward mode accumulation}.
Alternatively, we could start from $\adj{e}$ and pull it back through the sequence to obtain $\adj{t} = \dot{e}$, which is called \emph{reverse mode accumulation}.

The chain rule teaches us that we can relate tangents and cotangents of different variables through the \emph{Jacobian operator}.
If we formally denote the Jacobian as $\partial_p g$, we can denote its action on the complex tangent vector $\dot{p}$ as $\dot{g} = \partial_p g \, \dot{p}$.
We see that propagating a tangent forward through the computational graph can be performed by applying the Jacobian to a tangent vector on its right, which is commonly referred to as a \emph{Jacobian-vector product} (JVP).
The concrete action of a JVP for a generic non-holomorphic map $g$ is given by
\begin{equation}
    \label{eq:jvp}
    \dot{g}^k = \sum_l (\partial_{p^l} g^k) \, \dot{p}^l + (\partial_{\bar{p}^l} g^k) \, \dot{\bar{p}}^l.
\end{equation}
From this expression, we observe that the Jacobian acts as a \emph{real-linear} map on the complex vector $\dot{p}$, meaning that it is linear with respect to linear combinations of different tangent vector only when using real expansion coefficients.

To figure out how to pull back adjoints through the computational graph, we can substitute the JVP \cref{eq:jvp} for $\dot{g}$ into $\re \langle \adj{g}, \dot{g} \rangle$ and manipulate the expression to establish the equivalence \cref{eq:directional_derivative_bis}, $\re\langle \adj{g}, \dot{g} \rangle = \re\langle \adj{g}, \partial_p g \; \dot{p} \rangle = \re\langle (\partial_p g)^\dagger \, \adj{g}, \dot{p} \rangle$.
We then recognize $\adj{p} = (\partial_p g)^\dagger\, \adj{g}$, with $(\partial_p g)^\dagger$ the adjoint of the real-linear Jacobian with respect to the real inner product.
This operation is commonly referred to as a \emph{vector-Jacobian product} (VJP).
The concrete action of the VJP is given by
\begin{equation}
    \label{eq:vjp}
    \adj{p}_l = \sum_k \left[\adj{g}_k \overline{(\partial_{p^l} g^k)} + \overline{\adj{g}_k} (\partial_{\bar{p}^l} g^k)\right].
\end{equation}
In deriving this expression, we have equated $\re\left(\overline{\adj{g}_k} (\partial_{\bar{p}^l} g^k) \dot{\bar{p}}^l\right)$ with $\re\left(\adj{g}_k \overline{(\partial_{\bar{p}^l} g^k)} \dot{p}^l\right)$, crucially requiring the presence of the real part in the expression \cref{eq:directional_derivative_bis}, which can ultimately be traced back to the assumption that the final cost function is real-valued.

The procedure of reverse-mode accumulation can then be summarized as follows.
To implement the derivative of a cost function $e$, which is itself a composition of different functions, one only needs to know:
\begin{enumerate}
    \item the order in which the functions were applied,
    \item at which value of the parameters the functions were evaluated,
    \item the VJP action $\adj{p} = (\partial_p g)^\dagger\,\adj{g}$ of every function used in the composition.
\end{enumerate}
Points 1.\ and 2.\ can be easily handled automatically by any off-the-shelf AD engine.
The VJP action in point 3.\ needs to be manually specified for any function in the sequence that does not have a standard implementation.

Having introduced the notions of JVP, VJP and reverse-mode accumulation, we can take a step back and phrase things a bit more informally.
In particular, we will drop all complex conjugates, and generalize our notation from simple vectors in Euclidean vector spaces to generic objects such as tensors or contraction environments.
This intuitive informal notation should never lead to inconsistencies, as long as we keep the proper definition \cref{eq:vjp} in mind when actually implementing any VJP action
\footnote{As an alternative to the approach outlined here, one could incorporate complex numbers into AD by treating all variables and their complex conjugates as separate degrees of freedom. This approach is entirely equivalent, and can even be implemented by only tracking the adjoint variables as defined above.}.
We will also informally denote the VJP action $\adj{p} = (\partial_p g)^\dagger\, \adj{g}$ in terms of the actual vector-Jacobian product $\adj{p} = \adj{g} \, \partial_p g $ as a slightly more intuitive notation in the context of conceptual discussions, where we view the action of the adjoint $ (\partial_p g)^\dagger$ as the left action of $ \partial_p g$ on a cotangent and we drop all $\dagger$ symbols from the notation.
This simplified informal notation is systematically used in the main text, as well as in the remainder of this Appendix.

\subsection{Variational PEPS optimization using AD}

Returning to the problem of variational PEPS optimization of \cref{sec:peps_optimization}, we can now discuss how to evaluate the gradient of the energy density \cref{eq:peps_energy} using reverse-mode accumulation.
The most straightforward way to evaluate the VJP $\adj{x} \, \partial_p x$ in \cref{eq:peps_energy} is to examine how the contraction environment is computed in practice.
A contraction algorithm computes the environment through an iterative procedure in which an iterating function $f$ is applied to an initial guess $x_0$ until convergence is reached, $x^* = f(x^*, p)$.
We call this converged environment $x^*$ the \emph{fixed point} of $f$.
This gives rise to a sequence of intermediate objects $x_i$ through the procedure $x_1 = f(x_0, p)$, $x_2 = f(x_1, p)$, $\dots$, $x_{n-1} = f(x_{n-2}, p)$, and finally $x^* \equiv x_n = f(x_{n-1}, p)$, with $n$ the number of steps needed to reach convergence.
In this case, we can use the chain rule to write the Jacobian operator $\partial_p x$ in \cref{eq:peps_gradient_chain_rule} in terms of derivatives of $f$ to obtain
\begin{align}
    \label{eq:environment_jacobian_chain_rule_expanded}
    \adj{x} \, \partial_p &x = \nonumber \\
    &\adj{x} \sum_{k=1}^n \left(\left( \prod_{m=1}^{k-1} \partial_x f(x_{n-m}, p) \right) \partial_p f(x_{n-k}, p)\right).
\end{align}
As long as we make sure that an AD engine can compute the basic derivatives $\partial_x e$ and $\partial_p e$, as well as the VJP actions of $\partial_x f$, $\partial_p f$, we can use reverse-mode AD to automatically evaluate this expression by backpropagating the intermediate adjoints $\adj{x}_k$ through all $n$ iterations of $f$ according to \cref{eq:environment_jacobian_chain_rule_expanded}.

While this straightforward AD-based approach to PEPS optimization has been very successful, it has several drawbacks and limitations.
A first issue is that evaluating \cref{eq:environment_jacobian_chain_rule_expanded} requires different VJP actions evaluated in every iteration of the algorithm.
This means that all intermediate objects $x_k$ need to be stored in memory until the very end of the computation, leading to a significant memory overhead.
Second, this approach requires making the entire contraction algorithm differentiable.
While this is not a fundamental limitation, ensuring that every subroutine in the contraction algorithm is differentiable in a stable manner can be challenging and time consuming in practice.
In particular, the VJP action of certain primitive subroutines, such as singular value decompositions or eigensolvers, can suffer from instabilities that should be irrelevant to the overall energy gradient.
To ensure a correct energy gradient, these instabilities need to be addressed in the derivatives of the individual subroutines, often in a heuristic manner.

These issues can be resolved by directly evaluating the VJP action $\adj{x} \, \partial_p x$ using implicit differentiation.
The VJP action corresponding to commonly used contraction algorithms for two-dimensional tensor networks is constructed in \crefrange{sec:c4vctm}{sec:ctmrg}.


\section{Differentiation of eigenvalue problems}
\label{sec:differentiating_eigenvalue_problems}

Consider the eigenvalue equation
\begin{equation}
A v = \lambda v
\end{equation}
where $\lambda$ is for example specified as an extremal eigenvalue, and which we assume non-degenerate for simplicity. Let us denote the corresponding left eigenvector as $w$, i.e.\ $A^\dagger w = \lambda w$, with $w = v$ if $A$ is self-adjoint. If we consider both $\lambda$ and $v$ as functions of $A$, we can try to compute the tangents $\dot{\lambda}$ and $\dot{v}$ as function of $\dot{A}$. The defining equation is given by
\begin{equation}
\dot{A}
 v + A \dot{v} = \dot{\lambda} v + \lambda \dot{v}.
\end{equation}
We thus obtain
\begin{equation}
\dot{\lambda} = w^\dagger \dot{A} v,
\end{equation}
but face the well-known problem that $\dot{v}$ is not uniquely determined, i.e.\ the linear system
\begin{equation}
(A - \lambda \mathbb{1}) \dot{v} = \dot{A} v - \dot{\lambda} v = -(\mathbb{1} - v w^\dagger) \dot{A} v
\end{equation}
is underdetermined. For a given solution $\dot{v}$, so is $\dot{v}^\prime = \dot{v} + \alpha v$ for arbitrary $\alpha \in \mathbb{C}$. This freedom originates from the fact that the defining eigenvalue equation does not define $v$ uniquely, since $v^\prime = \mu v$ is also a valid solution for every $\mu \in \mathbb{C}$.

Imposing the normalization $\lVert v \rVert = 1$ results in the additional tangent condition $\re(v^\dagger \dot{v}) =0$. However, the linear system still admits the freedom $\dot{v}^\prime = \dot{v} + i \gamma v$ for $\gamma \in \mathbb{R}$, associated with the phase freedom of $v$. Further extending the tangent condition to $v^\dagger \dot{v} = 0$ eliminates the freedom completely, i.e.\ it makes the linear problem for $(\dot{\lambda}, \dot{v})$ as a function of $\dot{A}$ well determined.

Now consider the alternative strategy where we only consider $v$ as a function of $A$, determined through the equation
\begin{equation}
A v - (v^\dagger A v) v = 0
\end{equation}
Clearly, $v$ still needs to be an eigenvector for this equation to be satisfied, with $\lambda = v^\dagger A v$ the associated eigenvalue. Furthermore, projecting this equation onto $v^\dagger$ shows that this equation also contains the normalization condition $v^\dagger v = 1$, provided $\lambda \neq 0$. However, for a given $v$, $v\prime = e^{i \theta} v$ with $\theta \in \mathbb{R}$ is also a solution.

The resulting linear problem for the tangent $\dot{v}$ is given by
\begin{align*}
\dot{A} v + A \dot{v} - (\dot{v}^\dagger A v) v - (v^\dagger \dot{A} v) v - (v^\dagger A \dot{v}) v - \lambda \dot{v} = 0,
\end{align*}
which can be rewritten as
\begin{equation}
\big[(\mathbb{1}-v v^\dagger) A - \lambda \mathbb{1} \big] \dot{v} -\lambda v (\dot{v}^\dagger v) = -(\mathbb{1}-v v^\dagger) \dot{A} v.
\end{equation}
By projecting onto $v^\dagger$, we find that this equation indeed imposes $\re(v^\dagger \dot{v}) = 0$ as expected, but still admits $\dot{v}^\prime = \dot{v} + i \gamma v$ for $\gamma \in \mathbb{R}$ as a family of solutions.
However, when further restrictions on $v$ prohibit the phase freedom, e.g.~when $v$ takes the form of a Hermitian matrix as in the case of the corner matrix of $C_{4v}$, the resulting linear system for the (co)tangents has a unique solution.


\section{Boundary MPS contraction environments}
\label{sec:boundarymps_environments}

This Appendix contains explicit expressions for the evaluation of the PEPS energy density \cref{eq:peps_energy} using a boundary MPS contraction environment.
Here, we only explicitly consider the specific examples of evaluating horizontal and vertical nearest-neighbor two-body interactions $O_h$ and $O_v$ respectively.
Expectation values of other operators can be evaluated in an analogous way.

If the transfer operator $\cT$ of the PEPS norm network as defined in \cref{eq:boundarymps_transfer} has a Hermitian reflection symmetry, the bottom boundary MPS \cref{eq:boundarymps_mps_bot} is given by the Hermitian conjugate of the top boundary MPS \cref{eq:boundarymps_mps_top}.
In this case the tensors $x^t = (A_L^t, A_R^t, C^t)$ defining the top boundary MPS \cref{eq:boundarymps_mps_top} along with the corresponding left and right environments defined in \cref{eq:boundarymps_gl_top,eq:boundarymps_gr_top} form a complete \enquote{symmetric} contraction environment
\[(A_L^t, A_R^t, C^t, G_L^t, G_R^t)\]
that can be used to evaluate $O_h$ and $O_v$ as
\begin{equation}
    \label{eq:boundarymps_expval_symmetric}
    \diagram[1.0]{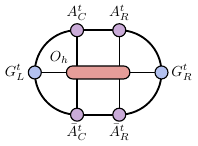},
    \qquad
    \diagram[1.0]{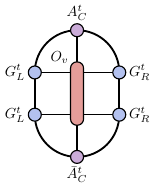}.
\end{equation}
Here, the auxiliary \enquote{center} tensor $A_C^t$ is defined as
\begin{equation}
    \label{eq:boundarymps_ac_top_definition}
    \diagram[1.0]{boundarymps_ac_top} =
    \frac{1}{2}\left(
        \diagram[1.0]{boundarymps_al_c_top} + \diagram[1.0]{boundarymps_c_ar_top}
    \right).
\end{equation}

If $\cT$ does not have a Hermitian reflection symmetry, computing expectation values additionally requires the bottom MPS tensors $x^b = (A_L^b, A_R^b, C^b)$ of \cref{eq:boundarymps_mps_bot} as well as the  \enquote{mixed} environment tensors of \cref{eq:boundarymps_gl_mixed,eq:boundarymps_gr_mixed}.
The corresponding \enquote{asymmetric} contraction environment
\[(A_L^t, A_R^t, C^t, A_L^b, A_R^b, C^b, G_L^t, G_R^t, G_L, G_R)\]
can be used to evaluate $O_h$ and $O_v$ as
\begin{equation}
    \label{eq:boundarymps_expval}
    \diagram[1.0]{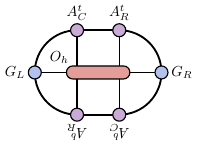},
    \qquad
    \diagram[1.0]{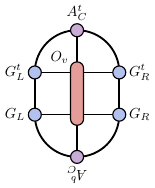}.
\end{equation}
Here the auxiliary \enquote{center} tensor $A_C^b$ is defined as
\begin{equation}
    \label{eq:boundarymps_ac_bot_definition}
    \diagram[1.0]{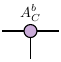} =
    \frac{1}{2}\left(
        \diagram[1.0]{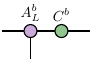} + \diagram[1.0]{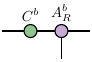}
    \right).
\end{equation}
Note that we have chosen not to use the bottom environment tensors $G_L^b$ and $G_R^b$, obtained from to the $(t \leftrightarrow b)$ version of \cref{eq:boundarymps_gl_top,eq:boundarymps_gr_top}, in the latter expression for evaluating the vertical nearest-neighbor interaction.
Making the opposite choice and not using $G_L^t$ and $G_R^t$ instead, or even using a combination of both approaches, would give the same results.

Backpropagating through the energy evaluation \cref{eq:boundarymps_expval_symmetric,eq:boundarymps_expval}, along with the definitions \cref{eq:boundarymps_ac_top_definition,eq:boundarymps_ac_bot_definition}, gives us adjoints of the environment tensors of the symmetric contraction environment
\[(\adj{A}_L^t, \adj{A}_R^t, \adj{C}^t, \adj{G}_L^t, \adj{G}_R^t)\]
and the asymmetric contraction environment
\[(\adj{A}_L^t, \adj{A}_R^t, \adj{C}^t, \adj{A}_L^b, \adj{A}_R^b, \adj{C}^b, \adj{G}_L^t, \adj{G}_R^t, \adj{G}_L, \adj{G}_R)\]
respectively.
Since we have chosen not to use the environments of the bottom MPS eigenvector in the contraction environment of \cref{eq:boundarymps_expval}, we can always set $\adj{G}_L^b = 0$ and $\adj{G}_R^b = 0$.
When using implicit differentiation, the adjoints $\adj{l}^\alpha$ and $\adj{r}^\alpha$ can be obtained from $\adj{A}_L^\alpha$ and $\adj{A}_R^\alpha$ by projecting out the corresponding null spaces $V_L^\alpha$ and $V_R^\alpha$ respectively, due to \cref{eq:boundarymps_al_top_parametrization,eq:boundarymps_ar_top_parametrization}.


\section{Mapping symmetric contraction environments}
\label{sec:symmetric_fixed_points}

In systems with a $C_{4v}$ point group symmetry, where the local PEPS tensor satisfies the symmetry conditions \cref{eq:peps_tensor_symmetries}, different contraction algorithms lead to contraction environments that can be interpreted as different equivalent realizations of the solution to an underlying variational problems \cite{vanderstraeten_variational_2022}.
In this Appendix we discuss how to map these different environments to each other explicitly.
In particular, we demonstrate how to obtain a corner tensor $C$, edge tensor $E$ and isometry $U$ that satisfy the characteristic equations \cref{eq:c4vctm_enlarged_corner_proj,eq:c4vctm_edge_proj}, as well as the additional gauge relation
\begin{equation}
    \label{eq:c4venv_gauge_relation}
    \diagram[1.0]{ctm_U} = \diagram[1.0]{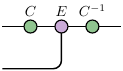}.
\end{equation}

We start by considering the $C_{4v}$ CTMRG algorithm as introduced in \cref{sec:c4vctm_contraction}.
After an appropriate gauge-fixing to ensure element-wise convergence, the $C_{4v}$ algorithm precisely returns a corner $C$, edge $E$ and isometry $U$ that satisfy \cref{eq:c4vctm_enlarged_corner_proj,eq:c4vctm_edge_proj}.
To additionally impose \cref{eq:c4venv_gauge_relation}, we can fix the global phase of the isometry $U$ such that
\begin{equation}
    \label{eq:c4venv_gauge_fixing}
    \tr\left(\diagram[1.0]{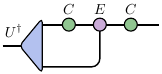}\right) = 1.
\end{equation}

However, aside from the procedure outlined in \cref{sec:c4vctm_contraction}, there are other CTMRG-like schemes that result in the same fixed point.
In particular, in the forward pass of \cref{alg:contract}, we could equivalently use
\begin{enumerate}
    \item The \texttt{eigh}-based $C_{4v}$ CTMRG algorithm as described in \cref{sec:c4vctm_contraction} \cite{nishino_corner_1996}.
    \item The QR-based $C_{4v}$ CTMRG algorithm as introduced in Ref.~\onlinecite{zhang_accelerating_2025a}.
    \item A Krylov-based contraction approach, known as the \emph{pulling-through} or \emph{fixed-point CTMRG} algorithm \cite{haegeman_diagonalizing_2017,fishman_faster_2018}, which is equivalent to the VUMPS algorithm in the presence of a $C_{4v}$ symmetry.
\end{enumerate}
Crucially, all of these forward schemes converge to the same fixed point, meaning they can use the same characteristic equation in the backwards pass.

To conclude this Appendix, we demonstrate how to transform a symmetric boundary MPS environment $(A_L^t, A_R^t, C^t, G_L^t, G_R^t)$ to an environment $(C, E, U)$ that is a root of the characteristic equations of \cref{sec:c4vctm_implicit_differentiation}.
First, we diagonalize the center gauge tensor by means of an SVD, $C^t = U^t \, S^t \, V^t$, and absorb the corresponding isometries $U^t$ and $V^t$ into the symmetric environment accordingly,
\begin{align}
    \diagram[1.0]{boundarymps_c_top} &\leftarrow \diagram[1.0]{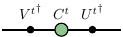}, \\
    \diagram[1.0]{boundarymps_al_top} &\leftarrow \diagram[1.0]{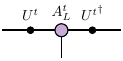}, \\
    \diagram[1.0]{boundarymps_ar_top} &\leftarrow \diagram[1.0]{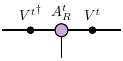}, \\
    \diagram[1.0]{boundarymps_left_fp_top} &\leftarrow \diagram[1.0]{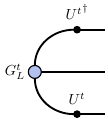}, \\
    \diagram[1.0]{boundarymps_right_fp_top} &\leftarrow \diagram[1.0]{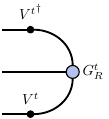}.
\end{align}
After this gauge transformation, in the presence of a $C_{4v}$ symmetry all tensors are all related by Hermitian conjugation up to a global phase.
By removing these global phases, we can impose the conditions
\begin{align}
    \diagram[1.0]{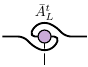} = \diagram[1.0]{boundarymps_ar_top}, \\
    \diagram[1.0]{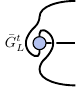} = \diagram[1.0]{boundarymps_left_fp_top}, \\
    \diagram[1.0]{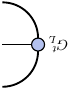} = \diagram[1.0]{boundarymps_right_fp_top}.
\end{align}
In this form, the environment tensor $G_L^t$ and the left isometry $A_L^t$ both define the same MPS.
Indeed, they only differ by a real-symmetric gauge transformation $X$ that satisfies $X^2 = C^t$,
\begin{equation}
    \diagram[1.0]{boundarymps_al_top} = \diagram[1.0]{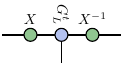}.
\end{equation}
The gauge matrix $X$ can be found as the leading eigenvector of the mixed $A_L^t-G_L^t$ transfer matrix,
\begin{equation}
    \diagram[1.0]{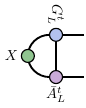} \propto \diagram[1.0]{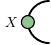}.
\end{equation}
Fixing the global phase of the eigenvector such that $X$ is real-symmetric, we obtain a symmetric environment satisfying \cref{eq:c4vctm_enlarged_corner_proj,eq:c4vctm_edge_proj,eq:c4venv_gauge_relation} as $(C, E, U) = (X, G_L^t, A_L^t)$.
Given this environment transformation, we can simply use the characteristic equations of \cref{sec:c4vctm_implicit_differentiation} to compute the energy gradient while using any boundary MPS algorithm \cite{zauner-stauber_variational_2018,vanhecke_tangentspace_2021,hauru_riemannian_2021} to perform the actual contraction.


\section{Alternative formulation of implicit differentiation for boundary MPS}
\label{sec:boundarymps_alternative}

In Section \ref{sec:boundarymps_implicit_differentiation} we presented optimality conditions $F(y,p)$ for PEPS contractions using boundary MPS, parametrized in terms of the 12 variables \[y = (l^t, r^t, C^t, G_L^t, G_R^t, l^b, r^b, C^b, G_L^b, G_R^b, G_L, G_R).\] 
Although evaluating the VJP of the contraction algorithm through a single linear problem is conceptually simple, it can become computationally inefficient as the environment bond dimension $\chi$ increases.

Here, we present an alternative approach where instead of solving one large linear problem, we decompose it into three smaller linear problems, two of which -- for top and bottom MPS -- can be solved independently and in parallel. This approach is summarized by the following alternative version of \cref{alg:energy}:

\begin{algorithm}[H]
    \caption{Alternative boundary MPS energy evaluation and gradient}
    \label{alg:energy_vumps}
    \begin{algorithmic}[1]
    \Require PEPS tensor $p$, Hamiltonian $H$, initial environment $x_0^t, x_0^b, x_0^g$ 

    \Function{energy\_alt}{$p, x_0, H$}
    \State $x^t \leftarrow \texttt{contract}(p, x_0^t)$ \Comment{Alg.~\ref{alg:contract} for top MPS}
    \State $x^b \leftarrow \texttt{contract}(p, x_0^b)$ \Comment{Alg.~\ref{alg:contract} for bottom MPS}
    \State $G_L, G_R \leftarrow \texttt{eig}(x^t, x^b, p, x^g_0)$ \Comment{\cref{eq:boundarymps_gl_mixed,eq:boundarymps_gr_mixed}}
    \State $x \leftarrow (x^t, x^b, G_L, G_R)$ \Comment{MPS contraction env.}
    \State \Return $e(p, x, H)$ \Comment{energy eval. using \cref{eq:boundarymps_expval}}
    \EndFunction

    \Statex
    \State $E,\frac{\td e}{\td p} \leftarrow \texttt{backpropagate}(\texttt{energy\_alt}, p, x_0, H)$
    \end{algorithmic}
\end{algorithm}

In this scheme, we split the \texttt{contract} routine for calculating the entire environment into 3 subroutines: one for the top boundary MPS $x^t$, the bottom boundary MPS $x^b$ and the mixed environment tensors $G_L,G_R$ (lines 2-4, respectively).
For the backward derivative we parametrize the top boundary MPS in exactly the same way as in \cref{sec:boundarymps_implicit_differentiation} with $y^t = (l^t, r^t, C^t, G_L^t, G_R^t)$ fulfilling optimality conditions $F^t(y^t,p)$, namely \crefrange{eq:boundarymps_l_top_proj}{eq:boundarymps_gr_top_bis}.
The same can be done in parallel for bottom MPS with input variables $y^b=(l^b, r^b, C^b, G_L^b, G_R^b)$ and characteristic equation $F^b(y^b,p)$. 

Mixed environment tensors $y^g = (G_L, G_R)$ are obtained as leading eigenvectors of a transfer matrix in \cref{eq:boundarymps_gl_mixed,eq:boundarymps_gr_mixed} that depends on $x^t$ and $x^b$, hence they are treated as functions $y^g(x^t,x^b,p)$. Therefore, in reverse mode AD, the engine must first evaluate the VJP actions of 
\begin{equation}
    \adj{x}^t=\adj{y}^g\partial_{x^t}y^g,\qquad \adj{x}^b=\adj{y}^g\partial_{x^b}y^g
\end{equation}
These contribute to the initial adjoints used for the backward pass through the top and bottom boundary MPS respectively. In practice, this requires first solving an additional linear problem for the leading eigenvalues and eigenvectors of the transfer matrix (see Ref. \onlinecite{Xie_AD_dominantEig_2020}).
We stress that it does not lead to a nested linear problem, but three smaller linear problems solved sequentially with possibility of parallelizing the calculation of derivatives of top and bottom MPS.


\section{CTMRG with unit cells}
\label{sec:asymmetric_ctmrg_unit_cells}

In this Appendix, we expand on \cref{sec:ctmrg_implicit_differentiation} by providing details on how to formulate the corresponding characteristic equations for non-trivial unit cells.
While most of the unit cell formulation follows straightforwardly from CTMRG without spatial symmetries on non-trivial unit cells~\cite{corboz_stripes_2011}, special care needs to be taken to assign the correct unit cell indices to the inverse singular values $s_\alpha$ as well as the modified square roots $s_\alpha^L$ and $s_\alpha^R$.

Recall that we label the spatial directions by $\alpha$, with $\alpha=1,2,3,4$ corresponding to the north, east, south and west directions, respectively.
We use matrix indexing to label the PEPS unit cell elements, where $r$ and $c$ denote the row and column indices,
\begin{equation}
    \label{eq:peps_nor_ctmrg_index}
    \diagram[1.0]{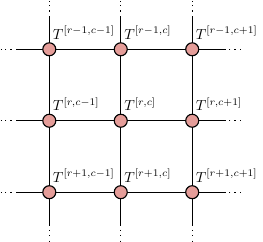}.
\end{equation}
Assuming four spatial directions and an $N_r \times N_c$-dimensional unit cell, the spatial index $d \in \mathbb{Z}_4$ and unit cell indices $r \in \mathbb{Z}_{N_r}$ and $c \in \mathbb{Z}_{N_c}$ obey modular arithmetic which will, however, not be denoted separately to reduce notational clutter.
With that in mind, we now provide the unit cell indices for \crefrange{eq:asymmctm_pl_tilde}{eq:asymmctm_c_e_tilde_fp} where the indices will be explicitly shown for all four spatial directions.
\begin{itemize}
    \item We first provide the indexing of the bond (inverse) singular values from \cref{eq:asymmctm_c_e_tilde_fp} which will appear in the subsequent diagrams:
    \begin{align}
        \label{eq:asymmctm_c_tilde_fp_index}
        \diagram[1.0]{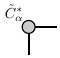} = \diagram[1.0]{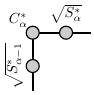}.
    \end{align}
    \begin{table}[H]
        \centering
        \begin{tabular}{ccccc}
        \toprule
        Tensor & $\alpha=1$ & $\alpha=2$ & $\alpha=3$ & $\alpha=4$ \\
        \midrule
        $\tilde{C}^*_\alpha$ & $(r,c)$ & $(r,c)$ & $(r,c)$ & $(r,c)$ \\
        $C^*_\alpha$ & $(r,c)$ & $(r,c)$ & $(r,c)$ & $(r,c)$ \\
        $\sqrt{S^*_{\alpha-1}}$ & $(r+1,c)$ & $(r,c-1)$ & $(r-1,c)$ & $(r,c+1)$ \\
        $\sqrt{S^*_\alpha}$ & $(r,c)$ & $(r,c)$ & $(r,c)$ & $(r,c)$ \\
        \bottomrule
        \end{tabular}
        \caption{Unit cell indices for \cref{eq:asymmctm_c_tilde_fp_index}.}
    \end{table}
    \begin{align}
        \label{eq:asymmctm_e_tilde_fp_index}
        \diagram[1.0]{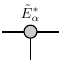} = \diagram[1.0]{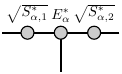}.
    \end{align}
    \begin{table}[H]
        \centering
        \begin{tabular}{ccccc}
        \toprule
        Tensor & $\alpha=1$ & $\alpha=2$ & $\alpha=3$ & $\alpha=4$ \\
        \midrule
        $\tilde{E}^*_\alpha$ & $(r,c)$ & $(r,c)$ & $(r,c)$ & $(r,c)$ \\
        $E^*_\alpha$ & $(r,c)$ & $(r,c)$ & $(r,c)$ & $(r,c)$ \\
        $\sqrt{S^*_{\alpha,1}}$ & $(r,c-1)$ & $(r-1,c)$ & $(r,c+1)$ & $(r+1,c)$ \\
        $\sqrt{S^*_{\alpha,2}}$ & $(r,c)$ & $(r,c)$ & $(r,c)$ & $(r,c)$ \\
        \bottomrule
        \end{tabular}
        \caption{Unit cell indices for \cref{eq:asymmctm_e_tilde_fp_index}.}
    \end{table}
    
    \item Next, we show the indices for the modified projectors from \crefrange{eq:asymmctm_pl_tilde}{eq:asymmctm_pr_tilde}. Note that the modified inverse square roots $s_\alpha^L$ and $s_\alpha^R$ appear in the open virtual legs of the half-infinite environment in \cref{eq:asymmctm_halfinf_env_svd} which needs to be considered when deriving their unit cell indices:
    \begin{align}
        \label{eq:asymmctm_pl_tilde_index}
        \diagram[1.0]{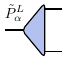} &= \diagram[1.0]{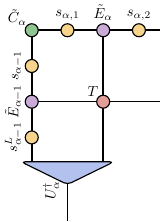}.
    \end{align}
    \begin{table}[H]
        \centering
        \begin{tabular}{ccccc}
        \toprule
        Tensor & $\alpha=1$ & $\alpha=2$ & $\alpha=3$ & $\alpha=4$ \\
        \midrule
        $\tilde{P}_\alpha^L$ & $(r,c)$ & $(r,c)$ & $(r,c)$ & $(r,c)$ \\
        $T$ & $(r,c)$ & $(r,c)$ & $(r,c)$ & $(r,c)$ \\
        $U_\alpha^\dagger$ & $(r,c)$ & $(r,c)$ & $(r,c)$ & $(r,c)$ \\
        $\tilde{C}_\alpha$ & $(r-1,c-1)$ & $(r-1,c+1)$ & $(r+1,c+1)$ & $(r+1,c-1)$ \\
        $\tilde{E}_{\alpha-1}$ & $(r,c-1)$ & $(r-1,c)$ & $(r,c+1)$ & $(r+1,c)$ \\
        $\tilde{E}_\alpha$ & $(r-1,c)$ & $(r,c+1)$ & $(r+1,c)$ & $(r,c-1)$ \\
        $s_{\alpha-1}^L$ & $(r+1,c-1)$ & $(r-1,c-1)$ & $(r-1,c+1)$ & $(r+1,c+1)$ \\
        $s_{\alpha-1}$ & $(r,c-1)$ & $(r-1,c)$ & $(r,c+1)$ & $(r+1,c)$ \\
        $s_{\alpha,1}$ & $(r-1,c-1)$ & $(r-1,c+1)$ & $(r+1,c+1)$ & $(r+1,c-1)$ \\
        $s_{\alpha,2}$ & $(r-1,c)$ & $(r,c+1)$ & $(r+1,c)$ & $(r,c-1)$ \\
        \bottomrule
        \end{tabular}
        \caption{Unit cell indices for \cref{eq:asymmctm_pl_tilde_index}.}
    \end{table}
    \begin{align}
        \label{eq:asymmctm_pr_tilde_index}
        \diagram[1.0]{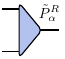} &= \diagram[1.0]{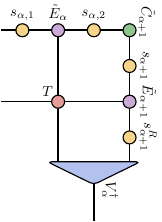}.
    \end{align}
    \begin{table}[H]
        \centering
        \begin{tabular}{ccccc}
        \toprule
        Tensor & $\alpha=1$ & $\alpha=2$ & $\alpha=3$ & $\alpha=4$ \\
        \midrule
        $\tilde{P}_\alpha^R$ & $(r,c)$ & $(r,c)$ & $(r,c)$ & $(r,c)$ \\
        $T$ & $(r,c+1)$ & $(r+1,c)$ & $(r,c-1)$ & $(r-1,c)$ \\
        $V_\alpha^\dagger$ & $(r,c)$ & $(r,c)$ & $(r,c)$ & $(r,c)$ \\
        $\tilde{C}_{\alpha+1}$ & $(r-1,c+2)$ & $(r+2,c+1)$ & $(r+1,c-2)$ & $(r-2,c-1)$ \\
        $\tilde{E}_\alpha$ & $(r-1,c+1)$ & $(r+1,c+1)$ & $(r+1,c-1)$ & $(r-1,c-1)$ \\
        $\tilde{E}_{\alpha+1}$ & $(r,c+2)$ & $(r+2,c)$ & $(r,c-2)$ & $(r-2,c)$ \\
        $s_{\alpha,1}$ & $(r-1,c)$ & $(r,c+1)$ & $(r+1,c)$ & $(r,c-1)$ \\
        $s_{\alpha,2}$ & $(r-1,c+1)$ & $(r+1,c+1)$ & $(r+1,c-1)$ & $(r-1,c-1)$ \\
        $s_{\alpha+1}$ & $(r-1,c+2)$ & $(r+2,c+1)$ & $(r+1,c-2)$ & $(r-2,c-1)$ \\
        $s_{\alpha+1}^R$ & $(r,c+2)$ & $(r+2,c)$ & $(r,c-2)$ & $(r-2,c)$ \\
        \bottomrule
        \end{tabular}
        \caption{Unit cell indices for \cref{eq:asymmctm_pr_tilde_index}.}
    \end{table}

    \item With these equations in mind, we find the following indices for the modified corner renormalization from \cref{eq:asymmctm_enlarged_corner_proj}:
    \begin{align}
        \label{eq:asymmctm_enlarged_corner_proj_index}
        \diagram[1.0]{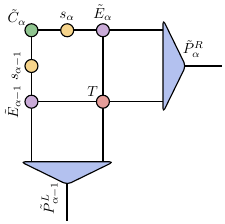} - \lambda_{C_\alpha} \diagram[1.0]{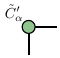} = 0.
    \end{align}
    \begin{table}[H]
        \centering
        \begin{tabular}{ccccc}
        \toprule
        Tensor & $\alpha=1$ & $\alpha=2$ & $\alpha=3$ & $\alpha=4$ \\
        \midrule
        $T$ & $(r,c)$ & $(r,c)$ & $(r,c)$ & $(r,c)$ \\
        $\tilde{P}_{\alpha-1}^L$ & $(r+1,c)$ & $(r,c-1)$ & $(r-1,c)$ & $(r,c+1)$ \\
        $\tilde{P}_\alpha^R$ & $(r,c)$ & $(r,c)$ & $(r,c)$ & $(r,c)$ \\
        $\tilde{C}_\alpha$ & $(r-1,c-1)$ & $(r-1,c+1)$ & $(r+1,c+1)$ & $(r+1,c-1)$ \\
        $\tilde{C}_\alpha^\prime$ & $(r,c)$ & $(r,c)$ & $(r,c)$ & $(r,c)$ \\
        $\tilde{E}_{\alpha-1}$ & $(r,c-1)$ & $(r-1,c)$ & $(r,c+1)$ & $(r+1,c)$ \\
        $\tilde{E}_\alpha$ & $(r-1,c)$ & $(r,c+1)$ & $(r+1,c)$ & $(r,c-1)$ \\
        $s_{\alpha-1}$ & $(r-1,c-1)$ & $(r-1,c+1)$ & $(r+1,c+1)$ & $(r+1,c-1)$ \\
        $s_\alpha$ & $(r,c-1)$ & $(r-1,c)$ & $(r,c+1)$ & $(r+1,c)$ \\
        \bottomrule
        \end{tabular}
        \caption{Unit cell indices for \cref{eq:asymmctm_enlarged_corner_proj_index}.}
    \end{table}

    \item The renormalization of the modified edges in \cref{eq:asymmctm_edge_proj} retains the same unit cell indices as in the standard CTMRG formulation:
    \begin{align}
        \label{eq:asymmctm_edge_proj_index}
        \diagram[1.0]{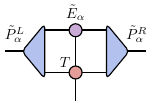} - \lambda_{E_\alpha} \diagram{asymmctm_e_tilde_relabel} = 0.
    \end{align}
    \begin{table}[H]
        \centering
        \begin{tabular}{ccccc}
        \toprule
        Tensor & $\alpha=1$ & $\alpha=2$ & $\alpha=3$ & $\alpha=4$ \\
        \midrule
        $T$ & $(r,c)$ & $(r,c)$ & $(r,c)$ & $(r,c)$ \\
        $\tilde{P}_\alpha^L$ & $(r,c-1)$ & $(r-1,c)$ & $(r,c+1)$ & $(r+1,c)$ \\
        $\tilde{P}_\alpha^R$ & $(r,c)$ & $(r,c)$ & $(r,c)$ & $(r,c)$ \\
        $\tilde{E}_\alpha$ & $(r-1,c)$ & $(r,c+1)$ & $(r+1,c)$ & $(r,c-1)$ \\
        $\tilde{E}_\alpha^\prime$ & $(r,c)$ & $(r,c)$ & $(r,c)$ & $(r,c)$ \\
        \bottomrule
        \end{tabular}
        \caption{Unit cell indices for \cref{eq:asymmctm_edge_proj_index}.}
    \end{table}

    \item Finally, the characteristic equations for the isometries and singular values as shown in \crefrange{eq:asymmctm_halfinf_env_proj_u}{eq:asymmctm_halfinf_env_proj_v} are described mostly by the same unit cell indices, which is why we only show the renormalization of $u$. The other two cases follow analogously:
    \begin{align}
        \label{eq:asymmctm_halfinf_env_proj_u_index}
        \diagram[1.0]{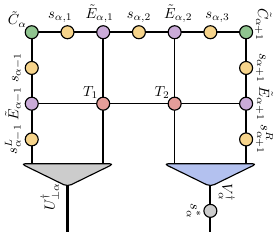} - \lambda_{S_\alpha} \diagram[1.0]{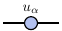} = 0.
    \end{align}
    \begin{table}[H]
        \centering
        \begin{tabular}{ccccc}
        \toprule
        Tensor & $\alpha=1$ & $\alpha=2$ & $\alpha=3$ & $\alpha=4$ \\
        \midrule
        $u_\alpha$ & $(r,c)$ & $(r,c)$ & $(r,c)$ & $(r,c)$ \\
        $T_1$ & $(r,c)$ & $(r,c)$ & $(r,c)$ & $(r,c)$ \\
        $T_2$ & $(r,c+1)$ & $(r+1,c)$ & $(r,c-1)$ & $(r-1,c)$ \\
        $U_{\perp\alpha}^\dagger$ & $(r,c)$ & $(r,c)$ & $(r,c)$ & $(r,c)$ \\
        $V_{\perp\alpha}^\dagger$ & $(r,c)$ & $(r,c)$ & $(r,c)$ & $(r,c)$ \\
        $\tilde{C}_\alpha$ & $(r-1,c-1)$ & $(r-1,c+1)$ & $(r+1,c+1)$ & $(r+1,c-1)$ \\
        $\tilde{C}_{\alpha+1}$ & $(r-1,c+2)$ & $(r+2,c+1)$ & $(r+1,c-2)$ & $(r-2,c-1)$ \\
        $\tilde{E}_{\alpha-1}$ & $(r,c-1)$ & $(r-1,c)$ & $(r,c+1)$ & $(r+1,c)$ \\
        $\tilde{E}_{\alpha,1}$ & $(r-1,c)$ & $(r,c+1)$ & $(r+1,c)$ & $(r,c-1)$ \\
        $\tilde{E}_{\alpha,2}$ & $(r-1,c+1)$ & $(r+1,c+1)$ & $(r+1,c-1)$ & $(r-1,c-1)$ \\
        $\tilde{E}_{\alpha+1}$ & $(r,c+2)$ & $(r+2,c)$ & $(r,c-2)$ & $(r-2,c)$ \\
        $s^L_{\alpha-1}$ & $(r+1,c-1)$ & $(r-1,c-1)$ & $(r-1,c+1)$ & $(r+1,c+1)$ \\
        $s_{\alpha-1}$ & $(r,c-1)$ & $(r-1,c)$ & $(r,c+1)$ & $(r+1,c)$ \\
        $s_{\alpha,1}$ & $(r-1,c-1)$ & $(r-1,c+1)$ & $(r+1,c+1)$ & $(r+1,c-1)$ \\
        $s_{\alpha,2}$ & $(r-1,c)$ & $(r,c+1)$ & $(r+1,c)$ & $(r,c-1)$ \\
        $s_{\alpha,3}$ & $(r-1,c+1)$ & $(r+1,c+1)$ & $(r+1,c-1)$ & $(r-1,c-1)$ \\
        $s_{\alpha+1}$ & $(r-1,c+2)$ & $(r+2,c+1)$ & $(r+1,c-2)$ & $(r-2,c-1)$ \\
        $s_{\alpha+1}^R$ & $(r,c+2)$ & $(r+2,c)$ & $(r,c-2)$ & $(r-2,c)$ \\
        $s_\alpha^*$ & $(r,c)$ & $(r,c)$ & $(r,c)$ & $(r,c)$ \\
        \bottomrule
        \end{tabular}
        \caption{Unit cell indices for \cref{eq:asymmctm_halfinf_env_proj_u_index}.}
    \end{table}
\end{itemize}


\bibliography{main}

\end{document}